\documentclass[preprint,showpacs,preprintnumbers,superscriptaddress]{revtex4-1}
\usepackage{amsmath,amssymb,graphicx,bm,float,siunitx,color}
\usepackage{natbib}
\usepackage{bm,float,siunitx,graphicx}
\usepackage[figuresright]{rotating}
\usepackage{color}
\usepackage{epstopdf}
\usepackage{mathrsfs}
\usepackage{subcaption}
\usepackage{float} 
\usepackage{overpic}% 为了使用 [H] 强制定位
\usepackage[font=small,labelfont=bf,justification=raggedright]{caption}

\newcommand{\be}{\begin{equation}}
\newcommand{\ee}{\end{equation}}

\def\[{\left[}
\def\]{\right]}

\newcommand{\m}{\mu}
\newcommand{\n}{\nu}

\begin{document}
\title{\boldmath Shadow of rotating black hole surrounded by dark matter
%Dark matter’s imprint on rotating black hole shadows
} %immersed in Plasma environment}
\author{Haiyuan Feng\footnote{Corresponding author}}
\email{Email address:  fenghaiyuan@sxnu.edu.cn}
\affiliation{School of Physics and Electronic Engineering, Shanxi Normal University, Taiyuan 030031, China}

\author{Ziqiang Cai}
\email{Email address: gs.zqcai24@gzu.edu.cn}
\affiliation{College of Physics, Guizhou University, Guiyang 550025, China}

\author{Rong-Jia Yang}
\email{Email address: yangrongjia@tsinghua.org.cn}
\affiliation{College of Physical Science and Technology, Hebei University, Baoding 071002, China}

\author{Jinjun Zhang\footnote{Corresponding author}}
\email{Email address: zhangjinjun@sxnu.edu.cn }
\affiliation{School of Physics and Electronic Engineering, Shanxi Normal University, Taiyuan 030031, China}

\begin{abstract}
Dark matter (DM), a fundamental cosmic component, motivates the study of its influence on black hole (BH) shadows, especially for spinning BHs confirmed by EHT observations. This work generalizes the Schwarzschild BH surrounded by DM to an axisymmetric Kerr BH using the Newman–Janis Algorithm (NJA), investigating the resulting event horizon and ergosphere structures. Employing null geodesics, we examine the effects of DM mass (\(\Delta M)\) on BH shadow, including its radius, distortion, and the associated energy emission rate. Our analysis reveals that DM has a negligible effect below a critical mass, once this threshold is surpassed, all BH structures expand significantly. Furthermore, DM robustly contributes to the shadow maintaining a near circular shape, even for highly spinning BHs. This pronounced structural expansion under high DM mass may potentially exceed current observational constraints, suggesting that DM must either be absent in the immediate vicinity of the BH or its localized mass must remain below this critical value to be consistent with astrophysical observations.
\end{abstract}

\maketitle

\section{Introduction}
Einstein’s general relativity (GR) fundamentally reshaped our view of spacetime and gravity, leading naturally to the prediction of black holes (BHs). Recent progress in gravitational experiments has ushered in a new stage in which electromagnetic observations can be used to probe the physical properties of ultra-compact objects such as BHs. Among the most significant achievements are the Event Horizon Telescope (EHT) measurements, which captured emission from the hot plasma circulating around the supermassive sources at the centers of M87 and the Milky Way \cite{1,2,3,4,5,6,7,8,9,10,11}, as well as infrared flare detections near the Galactic center by the gravity instrument \cite{Conroy:2023kec,GRAVITY:2023avo}. These developments provide an unprecedented opportunity to test BHs physics observationally, enabling detailed comparisons between high-resolution images and predictions obtained from General Relativistic Magneto-HydroDynamic (GRMHD) simulations of accreting plasma flows \cite{Johnson:2019ljv,EventHorizonTelescope:2020tst}.

It is remarkable that such high-precision observations now allow us to place stringent constraints on physically motivated BHs. In turn, this enables direct tests of GR and its possible extensions by searching for subtle departures from the Kerr geometry. The Kerr geometry represents the unique stationary and axisymmetric vacuum solution of Einstein field equations \cite{Mazur:2000pn}. Current observational breakthroughs are broadly consistent with this Kerr paradigm, which predicts that gravitational collapse in realistic astrophysical environments produces a rotating, electrically neutral BH \cite{Will:2014kxa,Yagi:2016jml,Penrose:1964wq,Kerr:1963ud}. As a result, the Kerr solution serves as the standard reference against which BH observations are interpreted, and any alternative spacetime must account for potential deviations from this framework.

In realistic astrophysical settings, the Kerr BH is not an isolated object but is immersed in a field of photons with diverse trajectories. Depending on their initial conditions, these photons may be captured by the event horizon, scatter back to infinity, or become temporarily confined to unstable circular orbits. The latter trajectories define the photon sphere, whose geometry delineates the boundary of the BH shadow. For an observer at large distance, the shadow-namely, the apparent dark silhouette-arises from the characteristic deflection of light in the strong-field region surrounding the BHs \cite{1983mtbh.book.....C}. A non-rotating BHs produces an exactly circular shadow, while rotation introduces a growing asymmetry in the silhouette, with the degree of distortion directly linked to the spin. The investigation of BH shadows began with Synge’s seminal analysis in the 1960s, where he derived the shadow cast by a spherically symmetric Schwarzschild BH \cite{Synge:1966okc}. Luminet later incorporated a surrounding accretion disk into this framework and provided a detailed computation of the resulting shadow size and appearance \cite{Luminet:1979nyg}. Subsequent extensions to rotating geometries demonstrated that the shadow of Kerr BH is no longer perfectly circular; instead, frame-dragging induces a characteristic asymmetry that encodes information about the spin \cite{1973CMaPh..31..161B}. These early developments laid the foundation for an extensive body of work exploring shadow structures in a wide variety of spacetime configurations. Recent studies have examined noncircular shadow features associated with alternative BH solutions \cite{I10,I11,I12,I13,I14,I16,I17,I18,I19,I20,I21} and horizonless compact objects such as wormholes \cite{I23,I24,I25,I26,I27,I28,I29,I30}, motivated by the possibility that shadow morphology may serve as a discriminating probe of gravitational dynamics beyond the standard Kerr paradigm.

Meanwhile, cosmological measurements indicate that nearly $85\%$ of the matter content of the Universe is composed of nonluminous dark matter (DM) \cite{WMAP:2010sfg}. Observations of stellar rotation curves—particularly the unexpectedly large orbital velocities of stars in the outer regions of galaxies—demonstrate that the visible galactic disk resides within an extended, approximately spherical DM halo \cite{Kafle:2014xfa,Battaglia:2005rj}. This unseen component dominates the galactic mass budget and plays a central role in shaping the gravitational potential on both galactic and cosmological scales.
Despite its overwhelming gravitational presence, the fundamental particle nature of DM remains fundamentally unknown. Numerous direct detection experiments most notably those targeting Weakly Interacting Massive Particles (WIMPs) have thus far yielded no conclusive evidence, challenging several of the leading particle physics candidates. While experiments such as the Cryogenic Dark Matter Search (CDMS) and DAMA/LIBRA have reported possible signals, these claims remain highly controversial, resulting in a mixed and often contradictory experimental landscape \cite{Bernabei:2018jrt,PICO:2017tgi,CRESST:2015txj,Bernabei:2013xsa,DAMA:2008jlt,Tsukamoto:2017fxq}. This persistent uncertainty further motivates astrophysical and gravitational probes, where the macroscopic effects of DM may offer complementary insights into its fundamental properties.

The advent of horizon scale BH imaging has opened a new avenue for investigating DM through its subtle imprints on BH observables. The presence of DM halo surrounding a compact object can modify the propagation of photons in its vicinity, leading to measurable departures in both the silhouette of the BH shadow and the weak gravitational deflection of light. In the context of EHT observations, such halos may enlarge the apparent shadow radius or introduce shape distortions that cannot be attributed solely to GR. Likewise, DM distributed near BHs can induce additional lensing effects, producing small but potentially detectable deviations in the trajectories of photons from background sources \cite{17,18,19,20,21,22,23,24,25,26,27,28,29,30,31,32,33,34,2022ApJ...935...91S,Chen:2025jay,Wan:2025gbm,Tsukamoto:2014tja,Nampalliwar:2021tyz}. These phenomena provide a complementary, indirect means of probing the properties of DM within strong gravity environments. In this paper, we extend the simplest DM model introduced in Ref. \cite{I13} to axisymmetric geometries by employing the Newman–Janis Algorithm (NJA). Within this framework, we aim to investigate the properties of the photon ring and the BH shadow, with particular emphasis on exploring how DM influences shadow related observables.

The parts of this paper are organized as follows. In Sec. II, we will introduce  Schwarzschild BH surrounded by DM and investigate the influence of DM mass on the event horizon, photon sphere, and shadow radius. In Sec. III, we will extend the Schwarzschild BH surrounded by DM to the axisymmetric case and further analyzed the event horizon and ergosphere structures of BH. Furthermore, we will study the null geodesic equations of Kerr BH surrounded by DM and examined the resulting shadow properties. By tracing photon trajectories, we will derive explicit expressions for the two impact parameters and present the corresponding shadow images in celestial coordinates. In addition, we will investigate the dependence of the shadow radius, distortion, and BH energy emission rate on the DM mass. We will also use the observational data of M87* and Sgr A* to place upper bounds on the allowed DM mass. Finally, we summarize the main results and conclusions.

%\section{Constraining dark matter mass}

% Observations and dynamical modeling indicate that the supermassive BH at the center of the Milky Way, Sgr A*, is embedded within a dense DM environment. In several galactic-halo models, the adiabatic growth of the central BH can generate a steep ‘spike’ in the surrounding DM distribution, producing a substantial enhancement of the density. This concentrated DM contributes to the gravitational field in the near horizon region and can therefore induce small but detectable modifications to photon trajectories, potentially imprinting observable deviations in the shadow size. Motivated by EHT observations of the supermassive BH Sgr A* \cite{EventHorizonTelescope:2022wkp}, the angular diameter of the BH shadow has emerged as a sensitive observational quantity for constraining theoretical models. In particular, the shadow size measured by the EHT provides a direct test of any modification to the spacetime geometry, including those induced by surrounding DM. In this section, we use the observed shadow diameter of Sgr A* to place bounds on the upper limit of DM mass. Our strategy is to compute the theoretical shadow diameter 
% $d_{\text{sh}}$ predicted by model and compare it with the EHT observational interval. The intersection between theory and measurement then delineates the viable region of parameter space, taking into account the full range of observational uncertainties.

\section{Schwarzschild black hole surrounded by dark matter}

%Within the framework of GR, numerous DM configurations have been proposed \cite{Moore:1999gc,Navarro:1994hi,Suto:1998xs,DiPaolo:2017geq}. 

The unusually high orbital velocities of stars in the outer regions of galaxies suggest that the luminous disks of these galaxies are embedded within a significantly larger, roughly spherical halo composed of DM. This DM halo, which envelops the central BH, may induce minor deviations from the standard spacetime geometry near the BH horizon. Moreover, the characteristic shape of the shadow image is influenced by the local environment around the BH, whereas the contour of the shadow is determined solely by the spacetime metric itself. Along this line of reasoning, BH solutions incorporating a DM halo and their impact on the BH shadow have emerged as a prominent area of research in recent years \cite{Moore:1999gc,Navarro:1994hi,Suto:1998xs,DiPaolo:2017geq,Hou:2018avu,Hou:2018bar,Haroon:2018ryd}. For instance, Figueiredo et al. developed asymptotically flat BH spacetimes sourced by generic anisotropic fluids and applied this framework to Navarro-Frenk-White (NFW), Hernquist, and Einasto profiles, with particular attention to geodesic observables and gravitational wave emission \cite{PhysRevD.107.104033}. Cardoso et al. constructed a fully relativistic BH in galaxy spacetime based on a Hernquist distribution \cite{PhysRevD.105.L061501}, whose light rings, gravitational lensing signatures, and shadow properties were subsequently investigated by Xavier et al. \cite{PhysRevD.107.064040}. More recently, Dehnen-type profiles, regarded as specific members of the broader Dehnen-Zhao double-power-law family, have been employed to study BH shadows, photon rings, innermost stable circular orbits, quasinormal modes, and constraints from the EHT \cite{Jha:2024ltc,Luo:2025xjb}. Beyond these profile-based models, Xu et al. derived static and rotating BH metrics for cold DM and Bose-Einstein-condensate halos \cite{Xu:2020jpv}; Haroon et al. examined the shadow and light deflection of a rotating BH surrounded by perfect fluid DM in the presence of a cosmological constant \cite{Haroon:2018ryd}; and Jusufi et al. constructed a rotating BH embedded in the M87 and universal rotation curve halo models, showing that realistic DM parameters lead to only negligible modifications of the M87* shadow \cite{Jusufi:2019nrn}. A key question is how the distribution of DM modifies the spacetime surrounding the galactic center BH, and a crucial aspect of this problem involves solving the Einstein field equations in the presence of DM. Several efforts in this direction have been reported in Refs. \cite{Cunha:2015yba,Hou:2018avu,Hou:2018bar,Haroon:2018ryd}; a related investigation concerning dark energy was carried out in Ref. \cite{Abdujabbarov:2015pqp}. Nevertheless, in all of these studies, some specific equation of state for DM or dark energy was presumed, rendering the outcomes highly dependent on the chosen model. Accordingly, we adopt the framework presented in Refs. \cite{I13,Saurabh:2020zqg}, which describes Schwarzschild BH embedded in a surrounding DM distribution.

The construction relies on two key assumptions. First, the DM is taken to be nonluminous and electromagnetically inert, allowing photons to traverse the medium without absorption. This assumption would break down for certain baryonic DM scenarios, where significant attenuation of light could occur and the interpretation would then depend on the detailed distribution and equation of state of the baryonic component. Second, the DM carries a finite mass density, which may be incorporated into the BH mass function as an effective additional mass. Although the model is fairly simplistic, it offers two advantages when describing the influence of DM on Schwarzschild BH. The model is characterized by independent fundamental parameters associated with the shell, namely the mass ratio \(\Delta M/M\), the inner boundary position \(r_s\), and the shell thickness \(\Delta r_s\). Additionally, it can be applied to any type of DM—not only the component that binds the entire galaxy, but also locally confined variants, such as non-luminous stars (brown dwarfs), stellar-mass BHs, primordial BHs, non-radiating dust, and others—all of which are gravitationally captured by the central supermassive BH.
Therefore, it is meaningful to discuss this model.

\begin{figure}[htbp]
\centering

% 第一行
\begin{minipage}{0.5\textwidth}
    \centering
    \includegraphics[width=\linewidth]{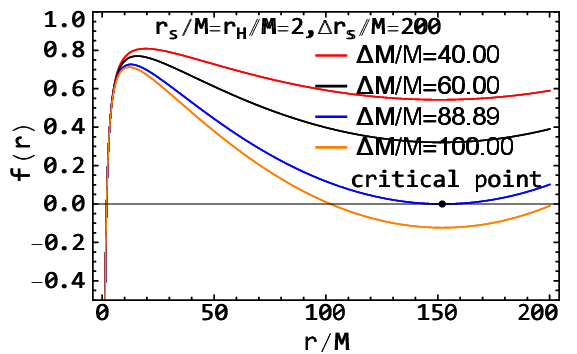}
\end{minipage}%
\hfill
\begin{minipage}{0.5\textwidth}
    \centering
    \includegraphics[width=\linewidth]{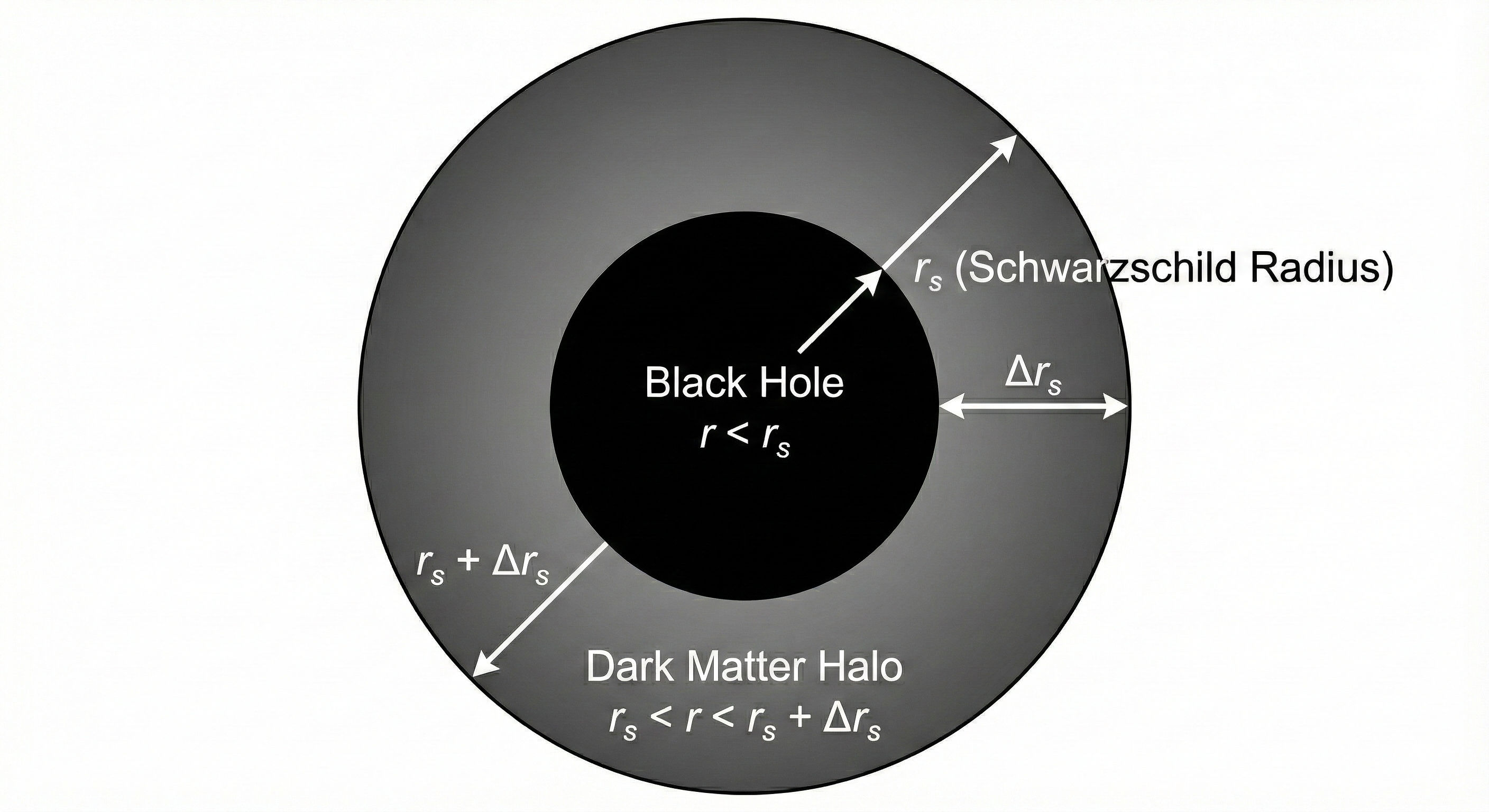}
\end{minipage}

\caption{\label{fig.1}{The left panel shows the corresponding radial dependence of $f(r)$, while the right panel illustrates the radial distribution of DM surrounding the Schwarzschild BH. In these plots, we fix the parameters to \( r_{s}/M = 2 \) and \( \Delta r_{s}/M = 200 \).}}
\end{figure}

The Schwarzschild BH surrounded by DM can be written as \cite{I13}
\be
\label{1}
d s^2=-f(r) d t^2+\frac{d r^2}{f(r)}+r^2\left(d \theta^2+\sin ^2 \theta d \phi^2\right),
\ee
with
\be
\label{2}
f(r)=1-\frac{2m(r)}{r},
\ee
where $m(r)$ denotes the mass function that incorporates the contribution from the surrounding DM. To account for the different physical regimes, the mass distribution is defined in a piecewise manner across three distinct radial domains, given by
\be
\label{3}
m(r)=\left\{\begin{array}{lr}
M, & r<r_s  \\
M+\Delta M\left(3-2 \frac{r-r_s}{\Delta r_s}\right)\left(\frac{r-r_s}{\Delta r_s}\right)^2, \quad
& r_s \leq r \leq r_s+\Delta r_s  \\
M+\Delta M, & r_s+\Delta r_s<r 
\end{array}\right.
\ee
where the functional $\left(3-2 \frac{r-r_s}{\Delta r_s}\right)\left(\frac{r-r_s}{\Delta r_s}\right)^2$ is constructed to ensure the continuity of both $m(r)$ and its derivative $m'(r)$. As illustrated on the right panel of Fig. \ref{fig.1}, the DM is predominantly concentrated within the interval $r_{s} \le r \le r_{s} + \Delta r_{s}$. %In this work, we set $r_{s} = r_{\rm H}$, identifying the inner boundary of the DM distribution with the BH horizon. 
Using the Einstein field equations $G_{\m\n}=8\pi T_{\m\n}$, one then obtain 
\be
\label{4}
T^t{ }_t=T^r{ }_r=-\frac{m^{\prime}(r)}{4 \pi r^2},
\ee
and
\be
\label{5}
T^\theta{ }_\theta=T^\phi{ }_\phi=-\frac{m^{\prime \prime}(r)}{8 \pi r} .
\ee

It is evident that both the density and pressure depend solely on the first and second derivatives of the mass function.
Since the chosen mass profile and its first derivative are continuous throughout the domain, whereas the second derivative exhibits only finite discontinuities at the boundaries, no Dirac-\(\delta\) terms arise. A detailed discussion is provided in Appendix A. Consequently, no additional singular contributions to the field equations or surface stress-energy tensors are generated at the interfaces. We also remark that the DM surrounding the BH may, in principle, fall into two categories: a configuration with positive energy density \(( \Delta M > 0 )\), corresponding to ordinary matter, and one with negative energy density \(( \Delta M < 0 )\). We will restrict our analysis to the physically motivated case \( \Delta M > 0 \).

Additionally, the left panel of Fig. \ref{fig.1} shows the radial dependence of the metric function \( f(r) \). We find that when the DM mass \( \Delta M \) becomes sufficiently large, its gravitational effect can shift the location of the Schwarzschild event horizon, thereby modifying the underlying BH structure. The metric function exhibits a critical point \(r_{*}\) at which the increasing gravitational influence of the DM distribution triggers the formation of an additional degenerate horizon. Imposing the conditions \(f(r_{*},\Delta M)=0\) and \(f'(r_{*},\Delta M)=0\), we obtain the critical values \(r_{*}/M=152\) and \(\Delta M/M=800/9\). For \(\Delta M/M<800/9\), the local minimum of \(f(r)\) outside the original BH horizon remains positive, so no additional horizon forms and the position of the event horizon is unchanged. As the DM mass increases, the enclosed mass and the effective compactness of the BH DM system increase, causing the metric function in the shell region to decrease. At the critical threshold, the local minimum of \(f(r)\) just reaches zero, corresponding to a degenerate horizon that separates configurations with and without an additional pair of horizons. Once \(\Delta M/M\) exceeds the critical value, this degenerate root splits into two distinct horizons. The outermost root then becomes the new event horizon, enclosing the central BH together with part of the DM shell, while the original BH horizon remains inside it. Therefore, the pronounced increase in the event horizon radius reflects the emergence of a new outer horizon branch associated with a sufficiently massive DM shell.

%Nevertheless, observational studies of the Galactic Center impose important constraints on any DM distribution surrounding Sgr A*. As noted in Ref. \cite{Kafle:2014xfa}, the total DM mass of the Milky Way is estimated to be \(M_{\rm DM} \simeq 6\times10^{11}–3\times10^{12},M_\odot\) \cite{Kafle:2014xfa}, whereas the central supermassive BH has a mass of only \(M_{\rm BH} \simeq 4.3\times10^{6}M_\odot\). In this situation, the presence of DM would inevitably distort the Schwarzschild radius and give rise to a shifted event horizon. Therefore, we reassess the horizon structure in the presence of DM to ensure that subsequent shadow constraints remain self-consistent.

% \begin{figure}[H]
% \centering
% \begin{minipage}{0.5\textwidth}
% \centering
% \includegraphics[scale=0.9,angle=0]{DMBH.eps}
% \end{minipage}%
% \begin{minipage}{0.56\textwidth}
% \centering
% \includegraphics[scale=0.1,angle=0]{DMBH.png}
% \end{minipage}%
% \caption{\label{fig.1}{ The left panel shows the corresponding radial dependence of $f(r)$, while the right panel illustrates the radial distribution of DM surrounding the Schwarzschild BH. In these plots, we fix the parameters to \( r_{s}/M = 2 \) and \( \Delta r_{s}/M = 200 \).}}
% \end{figure}

\subsection{Photon sphere radius and shadow structure of Schwarzschild black hole surrounded by dark matter}

For a generic static and spherically symmetric spacetime, the location of the photon sphere can be determined following the procedure outlined in Refs. \cite{I13,Pantig:2021zqe,Pantig:2020odu,Uniyal:2022vdu,Perlick:2021aok,Perlick:2015vta}. In such geometries, null geodesics admit an effective potential whose extremum corresponds to unstable circular photon orbits. By imposing the condition that the radial derivative of the effective potential vanishes, one obtains the photon sphere radius, which satisfies
\be
\label{6}
2 f(r_{\rm ph}) = r_{\rm ph} f'(r_{\rm ph}) .
\ee

This condition provides a general and coordinate independent criterion for identifying the photon sphere in any static, spherically symmetric metric. For the Schwarzschild BH surrounded by DM, as described by Eq. (1), previous studies have provided a detailed analysis of the associated photon sphere \cite{I13}. The main results relevant to our discussion can be summarized as follows. 

1: The DM is distributed such that the photon sphere lies between the event horizon \( r_{\rm H}/M=2\) and the inner boundary of DM layer \( r_{s} \). Under these conditions, photons propagate entirely through the vacuum Schwarzschild region, and the resulting shadow coincides with that of Schwarzschild BH. Accordingly, the photon sphere radius retains its standard value $r_{\mathrm{ph}} = 3M$. 

2:  Photon sphere located outside the DM shell $r_s + \Delta r_s < r_{\mathrm{ph}} < r_o$ . In such case, the effective mass governing photon orbits is simply the sum of the BH mass and the DM contribution, yielding $r_{\mathrm{ph}} = 3\bigl(M + \Delta M\bigr)$. 

3: Physically relevant configuration: photon sphere located within the DM layer. The only scenario of real interest arises when the innermost unstable photon orbit is embedded inside the DM distribution, such that $r_s < r_{\mathrm{ph}} < r_s + \Delta r_s$.
Solving the photon sphere condition with the metric functions leads to the following expression for the radius of the photon sphere:
\be
\label{7}
r_{\rm p h}=\frac{-\sqrt{F} \Delta \mathrm{r}_s+6 \Delta \mathrm{M} r_s\left(r_s+\Delta \mathrm{r}_s\right)+\Delta \mathrm{r}_s^3}{3 \Delta \mathrm{M}\left(2 r_s+\Delta \mathrm{r}_s\right)},
\ee
with
\be
\label{8}
\begin{aligned}
F\equiv & 12 \Delta M r_s \Delta r_s\left(\Delta r_s-3 M\right)-18 M \Delta M \Delta r_s^2+3 \Delta M r_s^2\left(4 \Delta r_s+3 \Delta M\right)+\Delta r_s^4.
\end{aligned}
\ee

%Regrettably, the scenario examined in this work involves DM contributions large enough to shift the Schwarzschild event horizon itself, rendering the three previously outlined cases inapplicable. In light of this, we turn to a numerical analysis to determine how the photon sphere radius responds to variations in DM parameters. 
Table 1 and 2 illustrate how the photon sphere location evolve as DM mass increases. Table 1 corresponds to the regime in which the DM mass is insufficient to modify the event horizon. In this case, increasing the mass leads to a slight reduction in the photon sphere radius. Table 2 corresponds to the regime in which the event horizon is modified, where the photon sphere radius increases by a significant factor. %A larger DM  contribution leads to pronounced changes in these characteristic radii, signaling significant alterations to the spacetime geometry. 
Physically, this trend is expected: the presence of additional matter in the vicinity of Schwarzschild BH effectively augments the total gravitational mass, thereby enlarging the fundamental length scales of the system.

Subsequently, following the notation introduced in Refs. \cite{Pantig:2021zqe,Pantig:2020odu,Uniyal:2022vdu,Perlick:2021aok,Perlick:2015vta}, we denote by \( r_{o} \) the radial position of the observer and by \( \alpha \) the angle measured with respect to the radial direction. Under these definitions, we obtain
\be
\label{9}
\cot\alpha
=\left.\frac{\sqrt{g_{rr}}}{\sqrt{g_{\phi\phi}}}
\frac{dr}{d\phi}\right|_{r=r_o}
=\left.\frac{1}{r\sqrt{f(r)}}\frac{dr}{d\phi}\right|_{r=r_o},
\ee
which links the geometric properties of photon trajectories to the shadow measured by a distant observer. Here we have restricted the photon motion to the equatorial plane \( \theta = \pi/2 \). With this choice, and by invoking elementary plane geometry considerations, one readily obtains the angular radius of the BH shadow. 
\be
\label{10}
\sin^2\left(\alpha_{\rm sh} \right)=\frac{h(r_{\rm ph})^2}{h(r_{o})^2},
\ee
where $h(r)$ is well-defined, $h(r)=r/\sqrt{f(r)}$. In the conventional limit where the observer is located at spatial infinity, \( r_{o} \rightarrow \infty \). In this case, the shadow radius can be derived
\be
\label{11}
R_{\rm sh}=r_{o}\sin{\alpha_{\rm sh}},
\ee
which serves as the principal physical quantity. Following the argument presented in Ref. \cite{I13}, one can make use of the standard property of spherical systems: only the matter enclosed within a given radius contributes to the gravitational force at that radius. Building on this idea, Ref. \cite{I13} proposes an approximate result
\be
\label{12}
R_{\rm sh}=r_{o}\sin{\alpha_{\rm sh}}\approx3\sqrt{3}\left( 
 M+\Delta \mathrm{M} -\frac{\Delta \mathrm{M}\left(-2 r_s+2 r_{\rm p h}+\Delta \mathrm{r}_s\right)\left(r_s-r_{\rm p h}+\Delta \mathrm{r}_s\right)^2}{\Delta \mathrm{r}_s^3}
 \right).
\ee

The essential point is that only the DM inside the photon sphere affects the shadow formation, while the exterior distribution can be neglected. From Tables 1 and 2, it is evident that the DM mass can indeed modify the BH shadow radius. In particular, once the mass exceeds the critical value, the shadow radius increases significantly, consistent with the strengthened gravitational field generated by the additional mass.

%Building on this conclusion, the next subsection uses the EHT measurements of Sgr A* to constrain the DM mass. Table 1 summarizes how the shadow radius responds to variations in the DM mass. As mass contribution increases, the shadow size expands monotonically, consistent with the strengthened gravitational field produced by the additional mass.

\begin{table*}[h]
\centering
\caption{\label{Tab.1} Table summarizes the dimensionless physical quantities for the  BH. We fix the parameters to \( r_{s}/M = 2 \) and \( \Delta r_{s}/M = 200 \).}
\setlength{\tabcolsep}{2pt} % 调整列间距，原默认为6pt
\renewcommand{\arraystretch}{1.05} % 调整行距，原默认为1.2
\begin{tabular}{|l|cccccc|}
\hline
\hline
BH & $\Delta M/M=20$ & $\Delta M/M=30$ & $\Delta M/M=40$& $\Delta M/M=50$ &
$\Delta M/M=60$ & $\Delta M/M=70$ \\
% \hline
% $r_{\rm H}/M$
% & 2 & 2 & 2 & 2 &2 & 2  \\
\hline
$r_{\rm ph}/M$
&2.99554 &2.99332&2.99111 & 2.98891 &2.98671  &2.98452 \\
\hline
$R_{\rm sh}/M$
&5.20392 &5.20780&5.21169 &5.21557 &5.21946 &5.22334   \\
\hline
\end{tabular}
\end{table*}

\begin{table*}[h]
\centering
\caption{\label{Tab.2} We present the variations of the relevant physical quantities in the regime where the DM mass is sufficiently large to modify the BH event horizon.}
\setlength{\tabcolsep}{2pt} % 调整列间距，原默认为6pt
\renewcommand{\arraystretch}{1.05} % 调整行距，原默认为1.2
\begin{tabular}{|l|cccccc|}
\hline
\hline
BH & $\Delta M/M=100$ & $\Delta M/M=120$ & $\Delta M/M=140$& $\Delta M/M=160$ &
$\Delta M/M=180$ & $\Delta M/M=200$ \\
\hline
$r_{\rm H}/M$
& 202 & 242 & 282 & 322 &362 & 402   \\
\hline
$r_{\rm ph}/M$
&303 &363&423 & 483 &543  &603 \\
\hline
$R_{\rm sh}/M$
&524.811 &628.734&732.657 &836.581 &940.504 &1044.430    \\
\hline
\end{tabular}
\end{table*}

\section{Rotating Kerr black hole surrounded by dark matter}
Considering rotating BHs is essential because astrophysical observations strongly suggest that most supermassive BHs possess significant angular momentum. Evidence from accretion disk spectroscopy, relativistic jet formation, and stellar orbital dynamics consistently indicates the presence of substantial spin, implying that a static BH description is insufficient. Moreover, the EHT images of M87* and Sgr A* reveal asymmetries in brightness and shadow structure that arise naturally from frame-dragging and other rotation-induced effects. Therefore, incorporating rotation is necessary for producing realistic photon trajectories and for achieving accurate comparisons between theoretical shadow predictions and observational data.

The Newman–Janis Algorithm (NJA) is a well-established procedure for generating rotating BH solutions from static metrics within the framework of GR \cite{B1,B2}. By employing a complex coordinate transformation together with a suitable tetrad representation, the NJA effectively introduces angular momentum into a non-rotating BH. This method has been widely used to construct rotating counterparts of various static solutions and provides a practical and physically motivated approach for exploring the geometry and observational signatures of spinning BHs \cite{B3,B4,B5,B6,B7}.

Following the conventional procedure, the first step is to rewrite the metric from Boyer–Lindquist (BL) coordinates into the Eddington–Finkelstein (EF) coordinates. This is achieved by applying an appropriate coordinate redefinition, which relates the time coordinate in BL form to the null coordinate \(u\) in EF form. Through this transformation, one obtains the EF coordinates required for implementing the subsequent steps of the NJA.

\be
\label{13}
d t=d u+\frac{d r}{f(r)}.
\ee
The metric then takes the following form 
\be
\label{14}
d s^2=-f(r) d u^2-2 d u d r+r^2 d \theta^2+r^2 \sin ^2 \theta d \phi^2 .
\ee
Using a null tetrad basis satisfying
\(l^\mu l_\mu = n^\mu n_\mu = m^\mu m_\mu = l^\mu m_\mu = n^\mu m_\mu = 0\) and
\(l^\mu n_\mu = -1, m^\mu \bar m_\mu = 1\), the corresponding contravariant metric  can be written as
\be
\label{15}
g^{\mu \nu}=-l^\mu n^\nu-l^\nu n^\mu+m^\mu \bar{m}^\nu+m^\nu \bar{m}^\mu,
\ee
with
\be
\label{16}
\begin{aligned}
l^{\m} & =\delta_r^{\m}, \\
n^{\m} & =\delta_u^{\m}-\frac{f(r)}{2} \delta_r^{\m}, \\
m^{\m} & =\frac{1}{\sqrt{2} r}\left(\delta_\theta^{\m}+\frac{i}{\sin \theta} \delta_\phi^{\m}\right),\\
\bar{m}^{\m} & =\frac{1}{\sqrt{2} r}\left(\delta_\theta^{\m}-\frac{i}{\sin \theta} \delta_\phi^{\m}\right).
\end{aligned}
\ee
Subsequently, we carry out the nontrivial coordinate transformations in the \((u, r)\) plane

\be
\label{17}
\begin{gathered}
u \rightarrow u-i a \cos \theta, \\
r \rightarrow r+i a \cos \theta,
\end{gathered}
\ee
where $a$ is the spin parameter. Meanwhile, we postulate that these coordinate transformations deform the original metric functions into new angular and spin dependent structures, namely \(f(r) \rightarrow F(r,a,\theta)\) and
\(r^2 \rightarrow \Sigma(r,a,\theta)=r^2+a^2\cos^2{\theta}\) (\(\lim _{a \rightarrow 0} F(r, a, \theta)=f(r), \lim _{a \rightarrow 0} \Sigma(r,a,\theta)=r^2 \)) \cite{Azreg-Ainou:2014aqa,Azreg-Ainou:2014pra}. Under the same procedure, the null tetrads are likewise promoted to their transformed counterparts.

\be
\label{18}
\begin{aligned}
l^{\m} & =\delta_r^{\m}, \\
n^{\m} & =\delta_u^{\m}-\frac{F}{2} \delta_r^{\m} ,\\
m^{\m} & =\frac{1}{\sqrt{2 \Sigma}}\left(\left(\delta_u^{\m}-\delta_r^{\m}\right) i a \sin \theta+\delta_\theta^{\m}+\frac{i}{\sin \theta} \delta_\phi^{\m}\right) ,\\
\bar{m}^{\m} & =\frac{1}{\sqrt{2 \Sigma}}\left(-\left(\delta_u^{\m}-\delta_r^{\m}\right) i a \sin \theta+\delta_\theta^{\m}-\frac{i}{\sin \theta} \delta_\phi^{\m}\right).
\end{aligned}
\ee
Furthermore, the rotating metric in the EF coordinates is 
\be
\label{19}
\begin{aligned}
\mathrm{d} s^2= & -F \mathrm{~d} u^2-2 \mathrm{~d} u \mathrm{~d} r+2 a\left(F-1\right) \sin ^2 \theta \mathrm{~d} u \mathrm{~d} \phi+\Sigma \mathrm{d} \theta^2 \\
& +2 a \sin ^2 \theta  \mathrm{~d} r \mathrm{~d} \phi+\sin ^2 \theta\left[\Sigma+a^2\left(2 -F\right) \sin ^2 \theta\right] \mathrm{d} \phi^2 .
\end{aligned}
\ee
The final stage of the NJA involves restoring the metric in Eq. (19) to BL form. This is accomplished by applying the following coordinate redefinitions, which eliminate the mixed terms and yield the standard BL structure.
\be
\label{20}
\mathrm{d} u=\mathrm{d} t+\lambda(r) \mathrm{d} r, \quad \mathrm{~d} \phi=\mathrm{d} \phi'+\chi(r) \mathrm{d} r,
\ee
with
\be
\label{21}
\begin{aligned}
  \lambda(r)&=-\frac{r^2+a^2}{r^2f(r)+a^2}, \\
\chi(r)& =-\frac{a}{r^2f(r)+a^2} .\\
\end{aligned}
\ee
Subsequently we obtain
\be
\label{22}
\begin{aligned}
F(r, a,\theta)&=\frac{r^2f(r)+a^2 \cos ^2 \theta}{\Sigma}.\\
\end{aligned}
\ee
Then, the rotating extension of the Schwarzschild BH metric in the presence of DM can be expressed as
\be
\label{23}
\begin{aligned}
d s^2&=-\left(1-\frac{2 \mathrm{Y} r}{\Sigma}\right) d t^2+\frac{\Sigma}{\Delta} d r^2+\Sigma d \theta^2-\frac{4ar\sin^2{\theta} \mathrm{Y} }{\Sigma} d t d \phi\\
&+\sin ^2 \theta\left[\frac{\left(r^2+a^2\right)^2-a^2 \Delta \sin ^2 \theta}{\Sigma}\right] d \phi^2
\end{aligned}
\ee
with
\be
\label{24}
\begin{aligned}
&\Delta(r,a) =r^2f(r)+a^2=r^2-2rm(r)+a^2,\\
&\mathrm{Y}(r) =\frac{r(1-f(r))}{2}=m(r),\\
&\Sigma(r,a,\theta)=r^2+a^2\cos^2{\theta},
\end{aligned}
\ee
where \(f(r)\) is specified by Eq. (2), and \(a\) denotes the BH spin parameter. It is worth noting that we have reinstated \(\phi'  \rightarrow\phi\). We observe that the solution reduces to the Kerr BH in the limit \(\Delta M = 0\). Furthermore, when the spin parameter is set to \(a = 0\), the spacetime smoothly recovers the spherically symmetric configuration described by Eq. (1). Moreover, any rotating solution generated through this procedure must be checked against the Einstein field equations. Fortunately, based on Ref. \cite{Jusufi:2019nrn}, we have already demonstrated in Appendix B that this class of metrics indeed satisfies the field equations, and the corresponding DM density and pressure profiles have been provided by

\be
\label{25}
\begin{aligned}
\rho&=-p_r=\frac{ \mathrm{m}^{\prime}(r) r^2}{4 \pi \Sigma^2},\\ p_\theta&=p_\phi=-p_r-\frac{\mathrm{m}^{\prime \prime}(r) r+2 \mathrm{m}^{\prime}(r)}{8 \pi \Sigma} .
\end{aligned}
\ee

Equations \eqref{25} describe the regular bulk energy density and pressures of the rotating DM configuration. The possible distributional sources at the two interfaces of the piecewise mass profile are examined separately in Appendix A, where we demonstrate that the corresponding surface stress-energy tensors vanish. With this assurance in place, we now proceed to analyze the physical properties of the rotating Kerr BH surrounded by DM.

\subsection{Event horizons}
Following the methodology used for the spherically symmetric case, we note that the horizon of an axisymmetric BH is likewise determined by \(\Delta(r,a)\). However, because the mass function is defined piecewise in our model, we obtain the relation
\be
\label{26}
\Delta(r,a)=\left\{\begin{array}{lr}
r^2+a^2-2rM, & r<r_s  \\
r^2+a^2-2rM-2r\Delta M\left(3-2 \frac{r-r_s}{\Delta r_s}\right)\left(\frac{r-r_s}{\Delta r_s}\right)^2, \quad
& r_s \leq r \leq r_s+\Delta r_s  \\
r^2+a^2-2rM-2r\Delta M. & r_s+\Delta r_s<r 
\end{array}\right.
\ee

The horizons are clearly governed by the spin parameter \(a\) as well as the DM mass parameters \(\Delta M\) and \(\Delta r_s\). A numerical examination of Eq. (26) shows that, for certain parameter choices, the equation admits two distinct solutions. These solutions correspond to the Cauchy horizon \(r_{-}\) (the smaller root) and the event horizon \(r_{+}\) (the larger root).

\begin{figure}[htbp]
\centering

% 第一行
\begin{minipage}{0.465\textwidth}
    \centering
    \includegraphics[width=\linewidth]{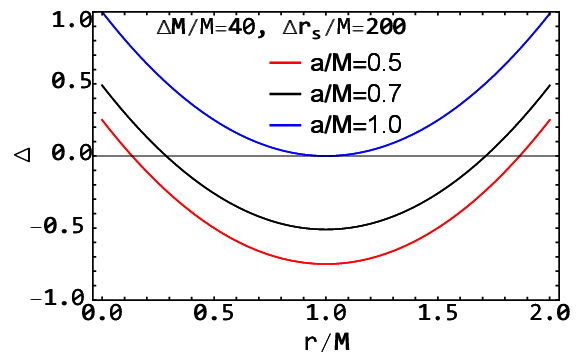}
\end{minipage}%
\hfill
\begin{minipage}{0.465\textwidth}
    \centering
    \includegraphics[width=\linewidth]{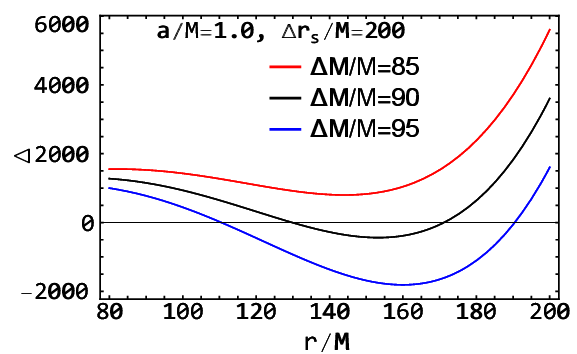}
\end{minipage}

\caption{\label{fig.2}{ The left panel illustrates the radial dependence of \(\Delta(r)\), where the parameters \(\Delta M/M=40\) and \(\Delta r_{s}/M=200\) are held fixed. The red, black, and blue curves correspond to spin parameters \(a=0.5\), \(a=0.7\), and \(a=1.0\), respectively. In the right panel, we set \(a=1.0\) and \(\Delta r_{s}/M=200\) to examine whether the DM mass influences the horizon structure. The red, black, and blue curves represent \(\Delta M/M=85\), \(\Delta M/M=90\), and \(\Delta M/M=95\), respectively.}}
\end{figure}

% \begin{figure}[H]
% \centering
% \begin{minipage}{0.5\textwidth}
% \centering
% \includegraphics[scale=0.9,angle=0]{delta1.eps}
% \end{minipage}%
% \begin{minipage}{0.56\textwidth}
% \centering
% \includegraphics[scale=0.9,angle=0]{delta2.eps}
% \end{minipage}%
% \caption{\label{fig.2}{ The left panel illustrates the radial dependence of \(\Delta(r)\), where the parameters \(\Delta M/M=40\) and \(\Delta r_{s}/M=200\) are held fixed. The red, black, and blue curves correspond to spin parameters \(a=0.5\), \(a=0.7\), and \(a=1.0\), respectively. In the right panel, we set \(a=1.0\) and \(\Delta r_{s}/M=200\) to examine whether the DM mass influences the horizon structure. The red, black, and blue curves represent \(\Delta M/M=85\), \(\Delta M/M=90\), and \(\Delta M/M=95\), respectively.}}
% \end{figure}

Figure 2 illustrates the dependence of \(\Delta(r)\) on the BH parameters. In the left panel, we analyze the effect of the spin \(a\) while assuming DM mass too small to alter the underlying horizon structure. In this regime, two distinct roots are obtained for \(|a/M|<1\), whereas in the extremal limit \(|a/M|\to 1\) the Cauchy and event horizons coalesce into a single degenerate root, mirroring the well-known behavior of the Kerr solution.

In contrast, the right panel of Fig. 2 examines the regime in which the DM mass becomes sufficiently large to modify the geometry. When the parameters \(a\) and \(\Delta r_{s}\) are fixed, the upper bound is determined by the conditions \(\Delta(r_{*})=0\) and \(\Delta'(r_{*})=0\). For the case \(a/M = 1\) and \(\Delta r_{s}/M = 200\), these equations yield \(r_{*}/M = 150.5\) and \(\Delta M/M = 88.3\). A critical upper limit for the DM mass emerges: the red curve, corresponding to a mass below this threshold, preserves the standard horizon configuration. However, the black and blue curves represent masses exceeding this limit, leading to a pronounced shift of the horizon radius, which increases by more than an order of magnitude. This substantial enlargement of the horizon closely parallels the trend identified in the spherically symmetric case. Increasing the DM mass at fixed spin and prescribed shell geometry strengthens the gravitational contribution of the shell until an additional horizon can appear at a much larger radius. In the rotating spacetime, however, the horizon locations are determined by $\Delta(r,a)=0$, and the rotational contribution modifies the condition for horizon formation. The onset of the additional horizon pair occurs when the outer local minimum of $\Delta$ first reaches zero. Thus, the simultaneous conditions $\Delta(r_*,a;\Delta M)=0$ and $\Delta'(r_*,a;\Delta M)=0$ identify a degenerate horizon separating configurations with different numbers of horizon roots. Above this threshold, the newly emerging outermost root defines the enlarged event horizon. Consequently, the critical DM mass generally depends on the spin parameter as well as the shell location and thickness, and must be determined separately for each rotating configuration. In our setup, the shell's inner boundary is tied to the outer horizon radius of the vacuum Kerr seed, so varying the spin also changes the shell location. The resulting shift in the critical mass therefore includes both the explicit rotational contribution to the horizon equation and the accompanying change in the mass distribution. The static threshold is recovered in the non-rotating limit with the corresponding shell parameters.

\subsection{Ergosphere}
The ergosphere is defined as the region enclosed between the event horizon \(r_+\) and the outer stationary limit surface. The stationary limit surface, also known as the infinite redshift surface, corresponds to the location where the metric component \(g_{tt}\) vanishes. Equivalently, it marks the boundary beyond which no observer can remain at rest relative to a distant inertial frame, due to the frame-dragging effect induced by the BH’s rotation. The inner and outer ergosurfaces are the two dimensional surfaces, which is 
\be
\label{27}
\left\{\begin{array}{lr}
r^2+a^2\cos^2{\theta}-2rM=0, & r<r_s  \\
r^2+a^2\cos^2{\theta}-2rM-2r\Delta M\left(3-2 \frac{r-r_s}{\Delta r_s}\right)\left(\frac{r-r_s}{\Delta r_s}\right)^2=0, \quad
& r_s \leq r \leq r_s+\Delta r_s  \\
r^2+a^2\cos^2{\theta}-2rM-2r\Delta M=0. & r_s+\Delta r_s<r 
\end{array}\right.
\ee

The ergosphere structure is similarly composed of three distinct regions. By setting \(r_{s} = r_{+}\) (The outer horizon of Kerr BH), only the latter two regions need to be considered, in line with the analysis of the horizon. When the DM mass is sufficiently large, it can shift the ergosphere’s location. Figure 3 illustrates these effects. The first row shows the ergosphere for spin parameters \(a/M = 0.5, 0.7, 1.0\), with \(\Delta M/M = 40\), \(\Delta r_{s}/M = 200\), and \(\theta = \pi/4\) and \(2\pi\). Across different values of \(a\), two distinct locations for the potential layers are observed. Only in the extremal case \((a/M = 1)\) at \(\theta = 2\pi\) does a single layer appear, which is consistent with the expected behavior of an extremal BH. The second row presents a particularly interesting scenario, in which the DM mass is sufficiently large to alter the ergosphere structure. We observe that for \(\Delta M/M = 90\) (black curve) and \(\Delta M/M = 95\) (blue curve), the contribution from the DM introduces additional roots in \(g_{tt}\). In other words, the mass becomes large enough to significantly shift the location and shape of the ergosphere, demonstrating the strong influence of surrounding matter on the BH.

Figure 4 shows the horizon and ergosphere regions plotted in the \((x, z)\) plane of Cartesian coordinate system centered on the BH. We take the largest and smallest roots as the outer and inner boundaries of horizon and ergosphere region, respectively. The two red curves denote the inner and outer boundaries of ergosphere, while the black curves represent inner and outer horizons. In the first row, we consider a relatively small DM mass, insufficient to modify the BH’s structure, and examine the effect of varying the spin parameter \(a\). As \(a\) approaches the extremal limit, the initially elliptical potential layers gradually separate into two nearly circular regions, while the horizons merge into a single surface. In the second row, larger DM masses generate new potential location and horizon regions, with the boundaries expanding substantially. In particular, in the last two panels, the red and black curves almost coincide, indicating that the geometric configuration has been completely altered. This behavior is physically consistent, as the increased DM mass enhances the gravitational field, thereby affecting the BH and modifying its spacetime structure.

\begin{figure}[htbp]
\centering

% 第一行
\begin{minipage}{0.465\textwidth}
    \centering
    \includegraphics[width=\linewidth]{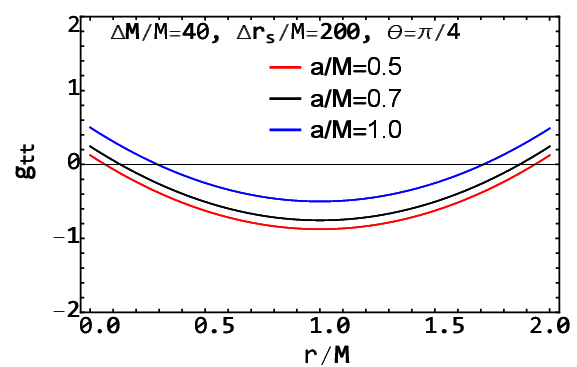}
\end{minipage}%
\hfill
\begin{minipage}{0.465\textwidth}
    \centering
    \includegraphics[width=\linewidth]{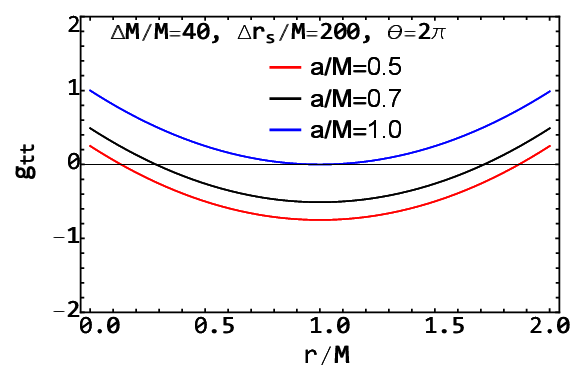}
\end{minipage}

\vspace{0.3cm}

% 第二行
\begin{minipage}{0.478\textwidth}
    \centering
    \includegraphics[width=\linewidth]{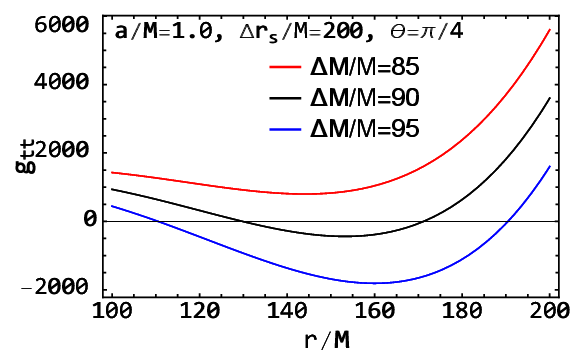}
\end{minipage}%
\hfill
\begin{minipage}{0.485\textwidth}
    \centering
    \includegraphics[width=\linewidth]{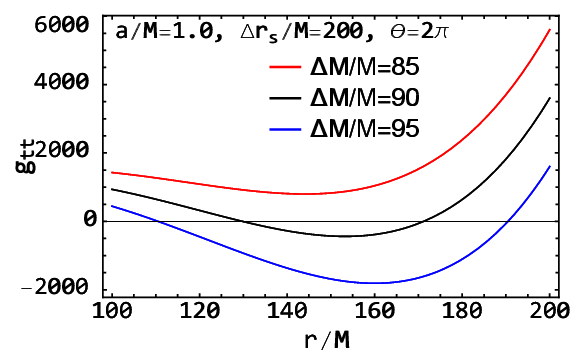}
\end{minipage}

\caption{\label{fig.3} The first row illustrates the variation of the ergosphere with the spin parameter, taking \(a/M=0.5,\,0.7,\) and \(1.0\). For both panels in this row, we fix \(\Delta M/M=40\) and \(\Delta r_s/M=200\), while considering \(\theta=\pi/4\) and \(2\pi\). For each curve, the zeros of the function determine the locations of the two corresponding potential layers. The second row shows the influence of the DM mass on the ergosphere locations. Similarly, for the two panels in the second row, we fix \(a/M=1\) and \(\Delta r_s/M=200\), and take \(\theta=\pi/4\) and \(2\pi\), respectively. We then investigate the cases \(\Delta M/M=85\), \(90\), and \(95\) to illustrate how the DM mass affects the locations of the ergosphere boundaries.}
\end{figure}

\begin{figure}[htbp]
\centering

% 第一行
\begin{minipage}{0.3\textwidth}
    \centering
    \includegraphics[scale=0.55]{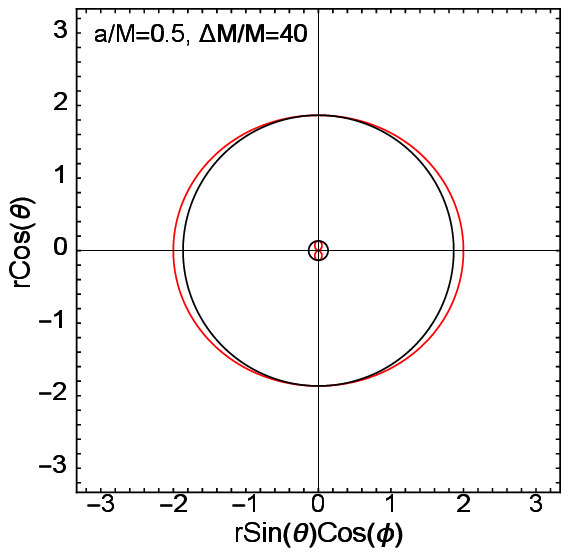}
\end{minipage}
\hspace{0.1cm}
\begin{minipage}{0.3\textwidth}
    \centering
    \includegraphics[scale=0.55]{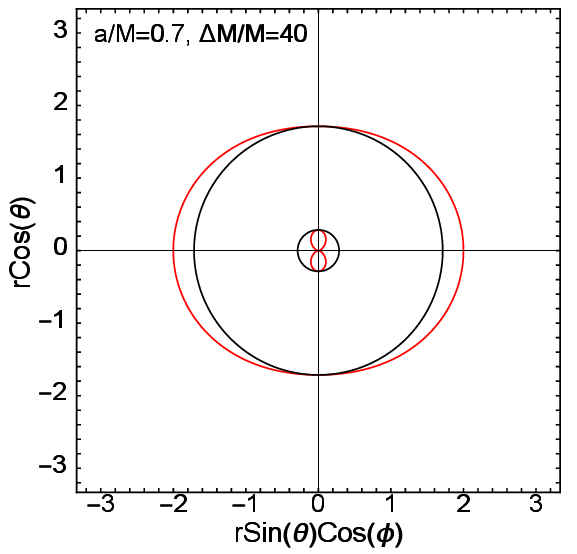}
\end{minipage}
\hspace{0.1cm}
\begin{minipage}{0.3\textwidth}
    \centering
    \includegraphics[scale=0.55]{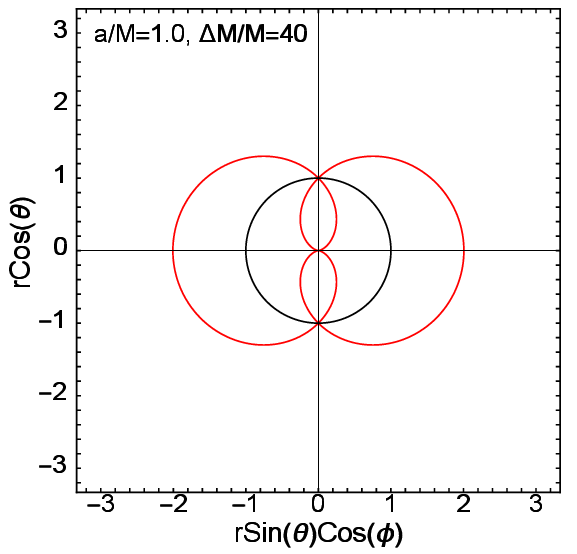}
\end{minipage}

\medskip

% 第二行
\begin{minipage}{0.3\textwidth}
    \centering
    \includegraphics[scale=0.55]{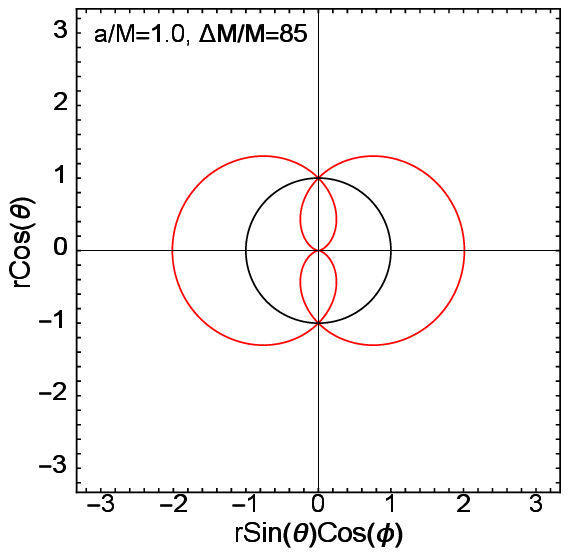}
\end{minipage}
\hspace{0.1cm}
\begin{minipage}{0.3\textwidth}
    \centering
    \includegraphics[scale=0.55]{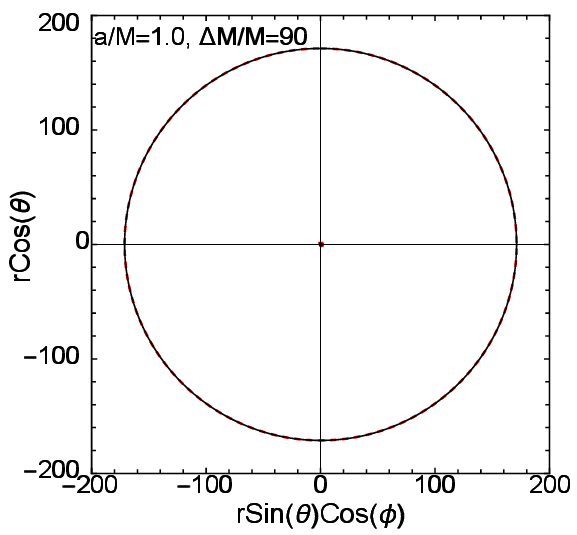}
\end{minipage}
\hspace{0.1cm}
\begin{minipage}{0.3\textwidth}
    \centering
    \includegraphics[scale=0.55]{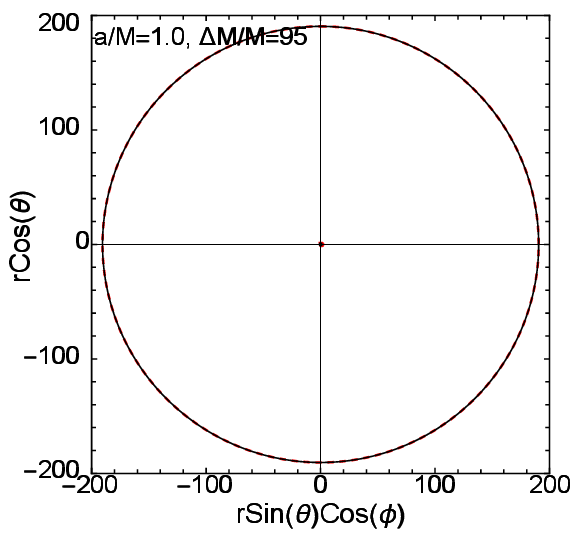}
\end{minipage}

\caption{\label{fig.4} In the \((x,z)\) plane of the Cartesian coordinate system, the ergospheres and horizons can be visualized more clearly. We adopt \(x=r\sin\theta\cos\phi\) and \(z=r\cos\theta\) as the horizontal and vertical coordinates, respectively. The first row shows the ergospheres and horizons for the spin parameters \(a/M=0.5,\,0.7,\) and \(1.0\), while fixing \(\Delta M/M=40\). The second row illustrates the cases \(\Delta M/M=85,\,90,\) and \(95\), with the spin parameter fixed at \(a/M=1\). In each panel, the outermost red curve denotes the outer boundary of the ergosphere, whereas the innermost red curve corresponds to the inner ergosphere boundary. Likewise, the innermost and outermost black curves indicate the inner and outer horizons, respectively.
}
\end{figure}

\section{Null geodesics and shadow cast}
One of the central objectives is to determine the shadow cast by the BH described by metric Eq. (23). In Sec. II, we examined the spherically symmetric configuration. In this section, we turn to the axisymmetric case. In contrast to the earlier results, the conclusions obtained here are expected to provide a more faithful representation of realistic astrophysical BH environments.
To achieve this, we begin by examining the null geodesics that govern photon motion in the specified spacetime geometry. The key idea underlying this approach is that the boundary of the shadow is dictated by the set of unstable photon trajectories: photons on these critical orbits either plunge into the BH or escape to infinity, and the transition between these behaviors defines the contour of the observable shadows. By identifying and analyzing these unstable circular photon orbits, we can therefore construct the precise shadow profile associated with the DM surrounded Kerr spacetime.

The equations governing photon trajectories can be derived efficiently through the Hamilton–Jacobi (HJ) formalism. Within this framework, the motion of a test particle is encoded in the HJ equation, which takes the form
\be
\label{28}
\frac{\partial S}{\partial \tau}=-\frac{1}{2} g^{\mu \nu} \frac{\partial S}{\partial x^\mu} \frac{\partial S}{\partial x^\nu},
\ee
here \(\tau\) denotes the affine parameter. For the rotating BH under consideration, the metric admits two Killing vector fields, \( \partial_t \) and \( \partial_\phi \), associated respectively with time-translation symmetry and axial symmetry. These geometric symmetries ensure the existence of two conserved quantities along the photon trajectory: the conserved energy \(E\) and the conserved angular momentum \(L\).
\be
\label{29}
\begin{aligned}
g_{tt}\dot{t} + g_{t\phi}\dot{\phi} &= -E, \\
g_{t\phi}\dot{t} + g_{\phi\phi}\dot{\phi} &= L.
\end{aligned}
\ee
These two conserved quantities play a central role in the subsequent analysis. Accordingly, we assume that the Jacobi action can be written in the form
\be
\label{30}
S=\frac{1}{2} m_0^2 \tau-E t+L \phi+S_r(r)+S_\theta(\theta),
\ee
where \(m_0\) denotes the particle’s rest mass. The functions \(S_r(r)\) and \(S_\theta(\theta)\) depend exclusively on the radial and polar coordinates, respectively, although their explicit forms remain to be determined. Inserting the ansatz for the Jacobi action (30) into the HJ equation (28), and using the fact that photons satisfy the massless condition, the equation separates into the following form
\be
\label{31}
\begin{aligned}
&S_r(r)=\int^r \frac{\sqrt{\mathcal{R}(r)}}{\Delta(r)} d r, \\
&S_\theta(\theta)=\int^\theta \sqrt{\Theta(\theta)} d \theta,
\end{aligned}
\ee
and
\be
\label{32}
\begin{aligned}
& \mathcal{R}(r)=\left[E\left(r^2+a^2\right)-a L\right]^2-\Delta\left(K+(L-a E)^2\right), \\
& \Theta(\theta)=K+\cos ^2 \theta\left(a^2 E^2-\frac{L^2}{\sin ^2 \theta}\right),
\end{aligned}
\ee
where \(K\) denotes the Carter constant. Because the HJ equation yields one separable relation for each coordinate, the system is completely integrable and therefore admits four independent constants of motion. For photon trajectories, the mass parameter is set to \(m_0 = 0\). By differentiating the HJ action with respect to these constants and imposing the condition that the resulting equations vanish, one obtains a set of first-order relations governing null geodesics. Solving this system leads directly to the standard form of the geodesic equations

\be
\label{33}
\begin{aligned}
\Sigma \frac{d t}{d \tau} & =\frac{r^2+a^2}{\Delta}\left[E\left(r^2+a^2\right)-a L\right]-a\left(a E \sin ^2 \theta-L\right), \\
\Sigma \frac{d r}{d \tau} & =\sqrt{\mathcal{R}(r)}, \\
\Sigma \frac{d \theta}{d \tau} & =\sqrt{\Theta(\theta)}, \\
\Sigma \frac{d \phi}{d \tau} & =\frac{a}{\Delta}\left[E\left(r^2+a^2\right)-a L\right]-\left(a E-\frac{L}{\sin ^2 \theta}\right).
\end{aligned}
\ee

The motion of photons in the rotating Kerr BH immersed in DM environment is entirely governed by the geodesic equations given in Eqs. (33). Of particular relevance for shadow formation are the unstable circular photon orbits, since these trajectories delineate the outermost paths from which light can still escape to a distant observer.
The contour of the BH shadow is therefore fixed by the conditions that define such marginally unstable null orbits, namely that the radial potential and its first derivative vanish. In other words, the photon must remain on a circular trajectory while being on the verge of radial instability. These requirements can be succinctly expressed as

\be
\label{34}
\left.\mathcal{R}(r)\right|_{r=r_{\rm p h}}=\left.\frac{d \mathcal{R}(r)}{d r}\right|_{r=r_{\rm p h}}=0,\left.\quad \frac{d^2 \mathcal{R}(r)}{d r^2}\right|_{r=r_{\rm ph}}>0,
\ee
where \(r_{\rm ph}\) represents the radius of the unstable circular photon orbits. To simplify the analysis, we introduce the dimensionless impact parameters \(\xi = L/E\) and \(\eta = K/E^2\), which characterize the photon’s angular momentum and the Carter constant normalized by its energy. Physically, \(\xi\) determines the trajectory’s azimuthal bending, while \(\eta\) encodes the motion in the polar direction. By substituting these definitions into Eq. (34) and solving the resulting system, one obtains explicit expressions for \(\xi\) and \(\eta\) that describe the critical photon orbits governing the boundary of BH shadow \cite{Yunusov:2024xzu,Jusufi:2019nrn,Hou:2018avu,Papnoi:2021rvw}.
\be
\label{35}
\begin{aligned}
& \left(r^2+a^2-a \xi\right)^2-\left[\eta+(\xi-a)^2\right]\left(r^2 f(r)+a^2\right)=0, \\
& 4 r\left(r^2+a^2-a \xi\right)-\left[\eta+(\xi-a)^2\right]\left(2 r f(r)+r^2 f^{\prime}(r)\right)=0.
\end{aligned}
\ee
At this point, Eq. (35) can be used to solve for
\be
\label{36}
\begin{aligned}
\xi & =\frac{\left(r^2+a^2\right)\left(r f^{\prime}(r)+2 f(r)\right)-4\left(r^2 f(r)+a^2\right)}{a\left(r f^{\prime}(r)+2 f(r)\right)}\\
&=\frac{r \left(a^2+r^2\right) \left(m'(r)+1\right)+\left(a^2-3 r^2\right) m(r)}{a \left(r m'(r)-r+m(r)\right)}, \\
\eta & =\frac{r^3\left(8 a^2 f^{\prime}(r)-r\left(r f^{\prime}(r)-2 f(r)\right)^2\right)}{a^2\left(r f^{\prime}(r)+2 f(r)\right)^2}\\
&=\frac{r^3 \left(-2 r  m'(r)\left(2 a^2-3 r m(r)+r^2\right)+m(r) \left(4 a^2-9 r m(r)+6 r^2\right)-r^3m'(r)^2-r^3\right)}{a^2 \left(r m'(r)-r+m(r)\right)^2} ,
\end{aligned}
\ee
here \(m(r)\) is mass function defined as in Eq. (3). Note that, in what follows, the radial coordinate \(r\) always refers to the photon sphere radius \(r_{\rm ph}\). We now proceed, in analogy with the spherically symmetric case, to classify the possible configurations of the photon sphere. Since we set \(r_s = r_{+}\), namely the outer event horizon of the Kerr BH, only the remaining two cases need to be considered. %However, by fixing \(\Delta r_{s}/M = 200\), the \(r_{\rm ph}\) is effectively confined within this range and does not extend beyond it. As a result, our discussion can be restricted to a single physically relevant configuration 

{\bf Case 1}: \( r_{s} < r_{\rm ph} < r_{s} + \Delta r_{s} \). In this regime, we obtain 
\be
\label{37}
\begin{aligned}
\xi&= \frac{\frac{2 r \left(a^2+r^2-2 r \Lambda (r)\right)}{r \left(\Lambda '(r)-1\right)+\Lambda (r)}+a^2+r^2}{a},\\
\eta&=\frac{r^3 \left(-2 r \left(2 a^2+r^2-3 r \Lambda (r)\right) \Lambda '(r)+\Lambda (r) \left(4 a^2+6 r^2-9 r \Lambda (r)\right)+r^3 \left(-\Lambda '(r)^2\right)-r^3\right)}{a^2 \left(r \left(\Lambda '(r)-1\right)+\Lambda (r)\right)^2},
\end{aligned}
\ee
with
\be
\label{38}
\begin{aligned}
&\Lambda (r)=M+\text{$\Delta $M} \left(3-\frac{2 (r-r_{+})}{\Delta r_{s}}\right) \left(\frac{r-r_{+}}{\Delta r_{s}}\right)^2,\\
&\Lambda'(r)=\frac{6 \text{$\Delta $M} \left(r-r_+\right) \left(r_+-r+\text{$\Delta $r}_s\right)}{\text{$\Delta $r}_s^3}.
\end{aligned}
\ee

{\bf Case 2}: \( r_{s}+ \Delta r_{s}< r_{\rm ph}\). In this region, we also derive
\be
\label{39}
\begin{aligned}
\xi&=\frac{a^2 (\text{$\Delta $M}+M+r)+r^2 (-3 \text{$\Delta $M}-3 M+r)}{a (\text{$\Delta $M}+M-r)},\\
\eta&=-\frac{r^3 \left(r (-3 \text{$\Delta $M}-3 M+r)^2-4 a^2 (\text{$\Delta $M}+M)\right)}{a^2 (\text{$\Delta $M}+M-r)^2}.
\end{aligned}
\ee

%Furthermore, we can directly obtain
% \be
% \label{37}
% \begin{aligned}
% \xi^2+\eta & =2 r^2+a^2+\frac{16\left(r^2 f(r)+a^2\right)}{\left(r f^{\prime}(r)+2 f(r)\right)^2}-\frac{8\left(r^2 f(r)+a^2\right)}{r f^{\prime}(r)+2 f(r)} \\
% & =\frac{\left(r m'(r)+m(r)\right) \left(r m'(r)\left(a^2+2 r^2\right)+ m(r)\left(a^2-6 r^2\right)+2 a^2 r\right)+r^2 \left(a^2+2 r^2\right)}{\left(r m'(r)-r+m(r)\right)^2}.
% \end{aligned}
% \ee

We now proceed to construct the shadow of Kerr BH surrounded by DM. The observer is assumed to be located at \((r_o, \theta_o)\), with \(r_o \to \infty\) and \(\theta_o\) representing the angular position on the observer’s sky. To characterize the apparent shape of the shadow, we introduce celestial coordinates \(\alpha\) and \(\beta\),

\be
\label{40}
\begin{aligned}
& \alpha=\lim _{r_o \rightarrow \infty}\left(-r_o^2 \sin \theta_o \frac{d \phi}{d r}\right), \\
& \beta=\lim _{r_o \rightarrow \infty}\left(r_o^2 \frac{d \theta}{d r}\right) ,
\end{aligned}
\ee
which correspond to the horizontal and vertical displacements of the photon as seen by the observer. These coordinates are expressed in terms of the conserved impact parameters \(\xi\) and \(\eta\), providing a direct connection between the photon trajectories in BH and the observable shadows profile \cite{Haroon:2018ryd,Yunusov:2024xzu,Jusufi:2019nrn,Hou:2018avu,Papnoi:2021rvw}.
By employing the null geodesic equations (33), one can express the celestial coordinates in terms of the photon’s conserved impact parameters as follows
\be
\label{41}
\begin{aligned}
\alpha & =-\frac{\xi}{\sin \theta_{o}}, \\
\beta & = \pm \sqrt{\eta+a^2 \cos ^2 \theta_{o}-\xi^2 \cot ^2 \theta_{o}}.
\end{aligned}
\ee
For an observer located in the equatorial plane \((\theta_0 = \pi/2)\), the expressions for the celestial coordinates simplify considerably, allowing a more straightforward mapping between the photon’s conserved quantities and the apparent shadows outline. 
\be
\label{42}
\begin{aligned}
& \alpha=-\xi, \\
& \beta= \pm \sqrt{\eta}.
\end{aligned}
\ee
This configuration is particularly useful for visualizing the shadows and comparing theoretical predictions with observations. The BH shadows can be constructed by generating a parametric curve in the celestial plane defined by the coordinates \(\alpha\) and \(\beta\), based on Eqs. (41) and (42). The curve is parametrized by the radius of the unstable circular photon orbits, \(r_{\rm ph}\). As previously discussed, the shadow corresponds to the region on the observer’s sky from which no photons arrive, since light rays within this zone are either captured by the BH or remain on unstable orbits. Consequently, the shape and size of the shadows are entirely determined by these critical photon trajectories.

Figure 5 shows the BH shadows in celestial coordinates for an observer at an inclination angle \(\theta_o = 90^\circ\). In the first row, the DM mass is fixed at \(\Delta M/M = 40, 60, 80\) to examine the effect of the spin parameter. As the spin increases, the shadow becomes progressively more distorted, with the deformation reaching its maximum in the extremal case, where a distinctive heart-shaped profile emerges.

In the second row, the spin parameter is fixed at \(a/M = 0.3, 0.7, 1.0\) to investigate the influence of the DM mass. For relatively small masses, the impact on the shadows remain weak, resulting only in a slight increase in its overall size. However, once the DM mass exceeds a critical value, the shadow undergoes a marked transformation, as shown in the third row. In this regime, the shadow radius becomes significantly larger and grows monotonically with increasing DM mass. These findings indicate that sufficiently large DM masses can strongly modify the BH shadows.

\begin{figure}[htbp]
\centering

% 第一行
\begin{minipage}{0.3\textwidth}
    \centering
    \includegraphics[scale=0.55]{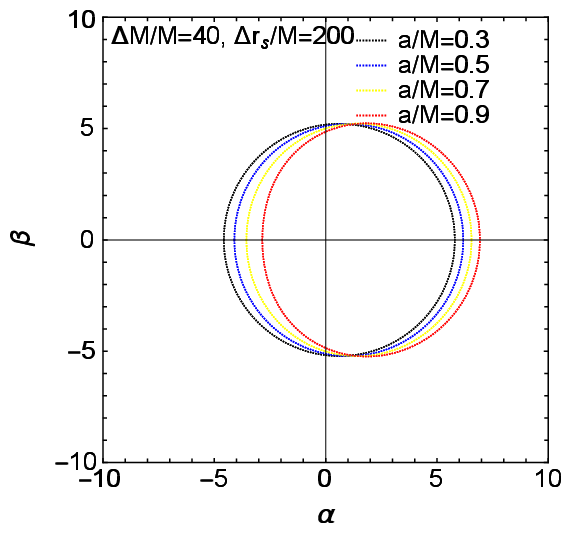}
\end{minipage}
\hspace{0.1cm}
\begin{minipage}{0.3\textwidth}
    \centering
    \includegraphics[scale=0.55]{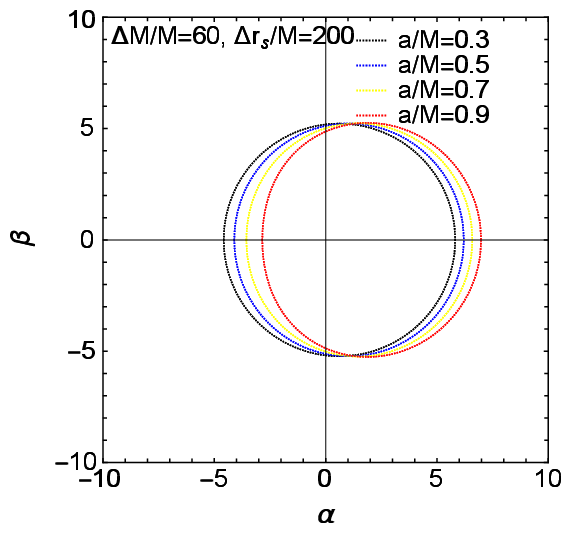}
\end{minipage}
\hspace{0.1cm}
\begin{minipage}{0.3\textwidth}
    \centering
    \includegraphics[scale=0.55]{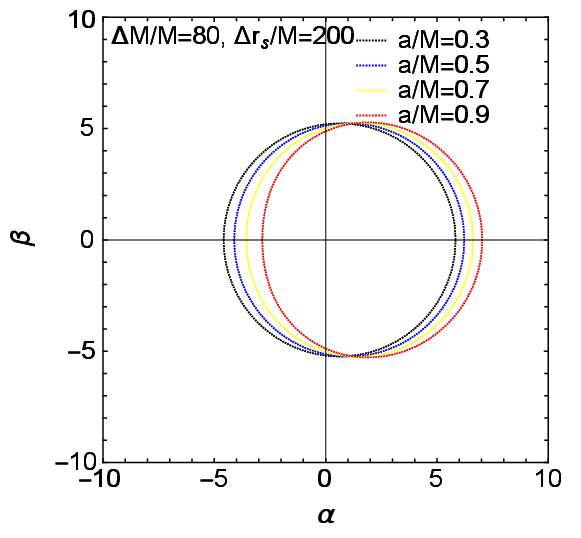}
\end{minipage}

\medskip

% 第二行
\begin{minipage}{0.3\textwidth}
    \centering
    \includegraphics[scale=0.55]{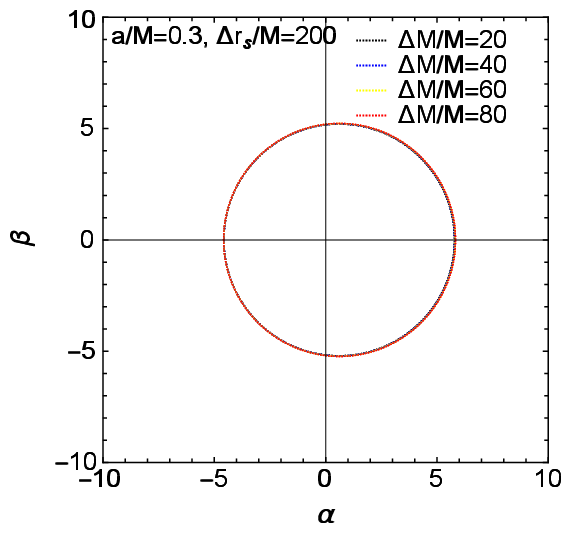}
\end{minipage}
\hspace{0.1cm}
\begin{minipage}{0.3\textwidth}
    \centering
    \includegraphics[scale=0.55]{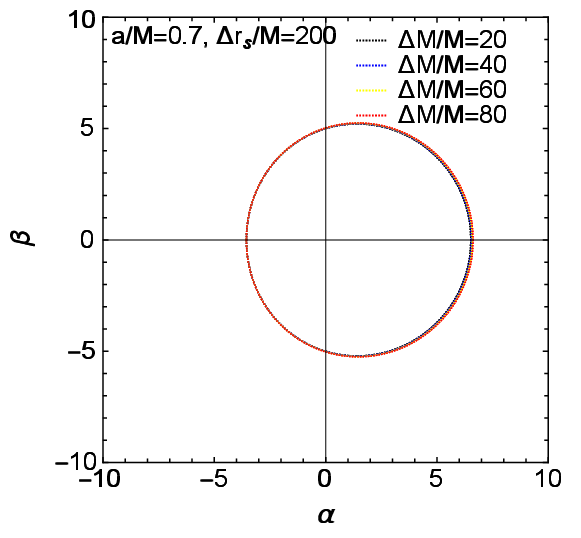}
\end{minipage}
\hspace{0.1cm}
\begin{minipage}{0.3\textwidth}
    \centering
    \includegraphics[scale=0.55]{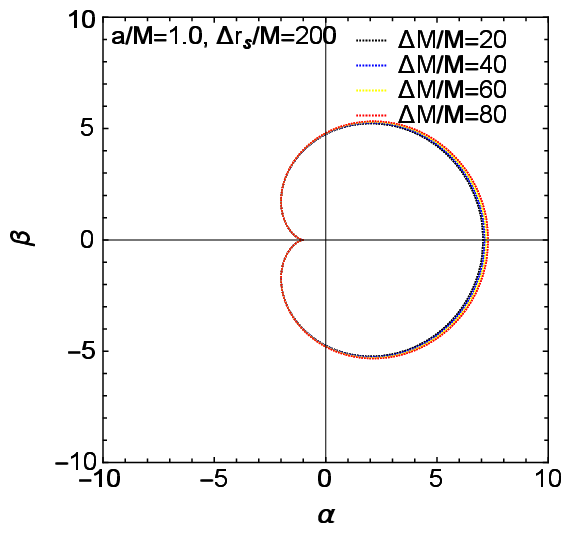}
\end{minipage}

% 第三行
\begin{minipage}{0.3\textwidth}
    \centering
    \includegraphics[scale=0.55]{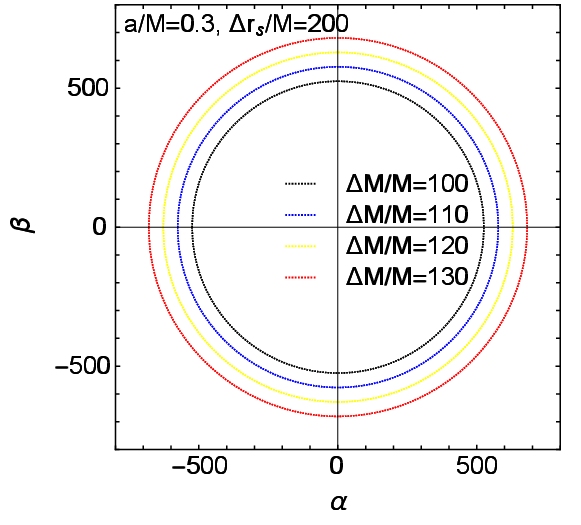}
\end{minipage}
\hspace{0.1cm}
\begin{minipage}{0.3\textwidth}
    \centering
    \includegraphics[scale=0.55]{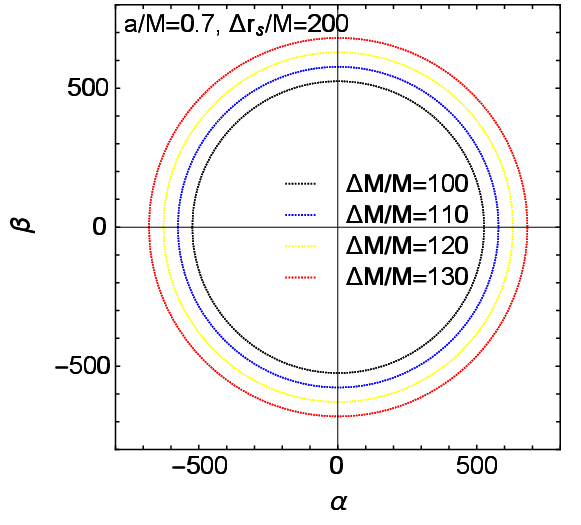}
\end{minipage}
\hspace{0.1cm}
\begin{minipage}{0.3\textwidth}
    \centering
    \includegraphics[scale=0.55]{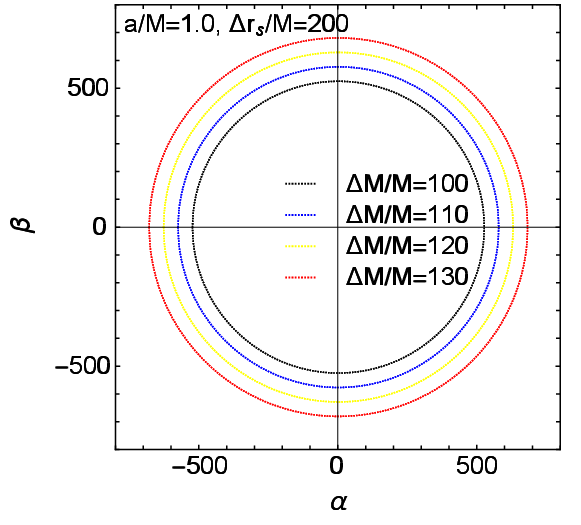}
\end{minipage}

\caption{\label{fig.5} BH shadow boundaries in the celestial plane $(\alpha,\beta)$ for an equatorial observer located at $\theta_o=90^\circ$. In all panels, the shell width is fixed at $\Delta r_s/M=200$. In the top row, the DM mass is fixed at $\Delta M/M=40$, $60$, and $80$ from left to right, respectively, while the four curves in each panel correspond to $a/M=0.3$, $0.5$, $0.7$, and $0.9$. This row therefore illustrates the effect of the spin parameter at fixed DM mass. In the middle row, the spin parameter is fixed at $a/M=0.3$, $0.7$, and $1.0$ from left to right, respectively, while the curves correspond to $\Delta M/M=20$, $40$, $60$, and $80$. In the bottom row, the columns retain the same fixed spin values, whereas the curves correspond to the larger DM masses $\Delta M/M=100$, $110$, $120$, and $130$. The middle and bottom rows thus illustrate the influence of the DM mass on the shadow morphology at fixed spin in the low and high mass regimes, respectively.}
\end{figure}

\subsection{Shadow observables}
The BH shadow provides a direct imprint of the underlying spacetime geometry, encoding information about both its structure and characteristic length scales. As a result, shadow observations offer a powerful tool for testing gravity in the strong-field regime and for constraining the parameters of BHs. To make quantitative use of this method, one must define observable quantities that capture the size and shape of the shadow. A pioneering framework was introduced by Hioki and Maeda \cite{Hioki:2009na,Schee:2008kz}, who proposed two key observables to characterize BH shadows. The first is the shadow radius, which measures the overall size of the silhouette. The second is the distortion parameter, designed to quantify departures from circularity. While a perfectly circular shadow is a hallmark of static BHs, rotation generally induces asymmetry and deformation. Accordingly, the distortion parameter provides a convenient means to assess the deviation of rotating BH shadows from the static case, with the linear shadow radius serving as a practical and directly measurable quantity \cite{Amir:2016cen,Hioki:2009na}.

As illustrated in Fig. 6, four characteristic points on the shadow boundary are introduced to define its geometry. These include the top point \((\alpha_t, \beta_t)\) and the bottom point \((\alpha_b, \beta_b)\), as well as the point \((\alpha_r, 0)\), which corresponds to photons on unstable retrograde circular orbits as seen by an observer in the equatorial plane. Similarly, the point \((\alpha_l, 0)\) is associated with photons following unstable prograde circular orbits. Based on these reference points, the shadow radius can be defined as
\be
\label{43}
R_{\rm sh}=\frac{\left(\alpha_t-\alpha_r\right)^2+\beta_t^2}{2\left(\alpha_r-\alpha_t\right)},
\ee
here we have made use of the shadow’s symmetry, which implies \(\alpha_b = \alpha_t\) and \(\beta_b = -\beta_t\). It is worth noting that when the shadow is perfectly circular, this definition reduces to that of spherically symmetric BH shadows. With these relations in place, we can now proceed to define
\be
\label{44}
\delta_s=\frac{|\bar{\alpha}_l-\alpha_l|}{R_{\rm sh}},
\ee
where \(\bar{\alpha}_l\) represents the horizontal coordinate of the leftmost point of the reference circle. The ratio of the difference between this coordinate and the leftmost point of BH shadow to the shadow radius provides a measure of how much the shadow departs from an ideal circular shape.

\begin{figure}[htbp]
\centering
\begin{minipage}{0.9\textwidth}
\centering
\includegraphics[scale=0.4,angle=0]{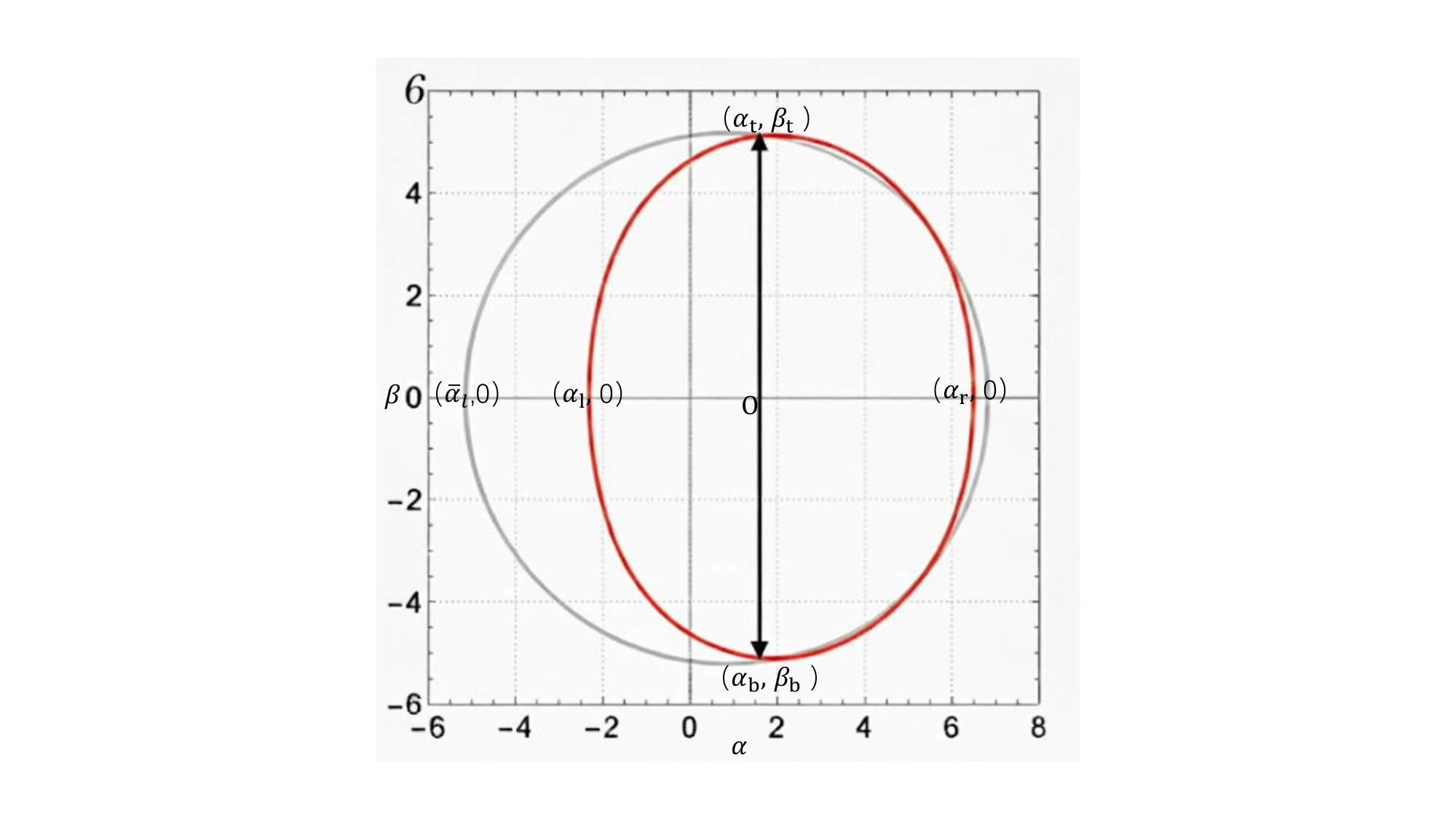}
\end{minipage}
\caption{\label{fig.6} A schematic illustration of the shadow cast by a rotating BH. In this diagram, the rightmost, leftmost, upper, and lower points on the shadow boundary are denoted by \((\alpha_r, 0)\), \((\alpha_l, 0)\), 
\((\alpha_t,\beta_t)\), and \((\alpha_b, \beta_b)\), respectively, with 
\(\alpha_t = \alpha_b\). In the non-rotating limit, the shadow is symmetric about the vertical axis, and the magnitudes of \(\alpha_r\) and \(\alpha_l\) become identical.}
\end{figure}

Figure 7 displays the numerical results for the dependence of the shadow radius and the distortion parameter on the DM mass and the spin. The first row illustrates the variation of the \(R_{\rm sh}\) with \(a\) and \(\Delta M\), while the second row shows the corresponding behavior of the distortion parameter.

From the first column, we observed that for DM masses below the critical value, an increase in the spin parameter leads to a larger shadow radius, and the distortion likewise becomes more pronounced as \(a\) approaches the extremal limit. In the second column, where the spin is held fixed, the shadow radius increases approximately linearly with \(\Delta M\), whereas the distortion decreases monotonically. This behavior indicates that the presence of DM tends to circularize the shadow.

\begin{figure}[H]
\centering

% 第一行
\begin{minipage}{0.31\textwidth}
    \centering
    \includegraphics[scale=0.53]{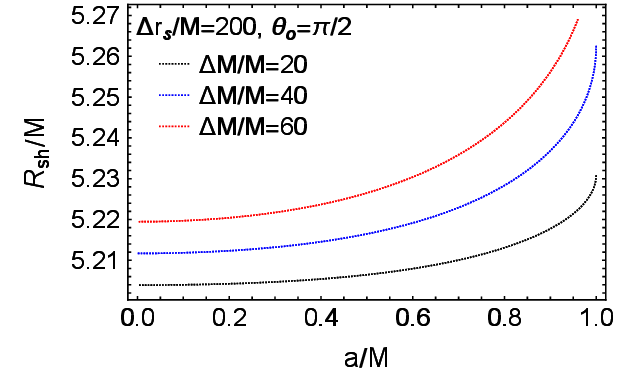}
\end{minipage}
\hspace{0.1cm}
\begin{minipage}{0.31\textwidth}
    \centering
    \includegraphics[scale=0.53]{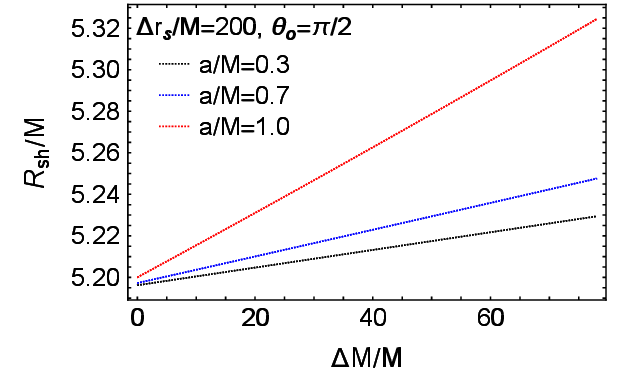}
\end{minipage}
\hspace{0.1cm}
\begin{minipage}{0.31\textwidth}
    \centering
    \includegraphics[scale=0.53]{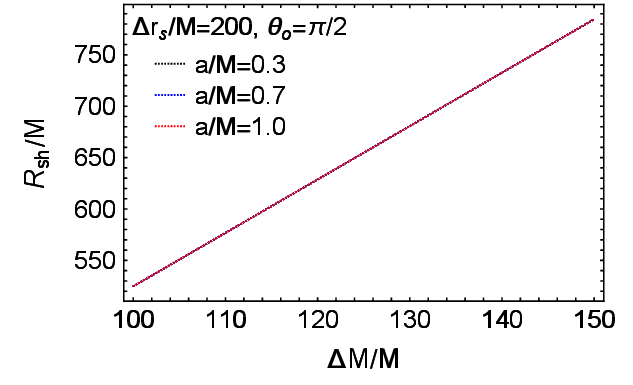}
\end{minipage}

\medskip

% 第二行
\begin{minipage}{0.31\textwidth}
    \centering
    \includegraphics[scale=0.53]{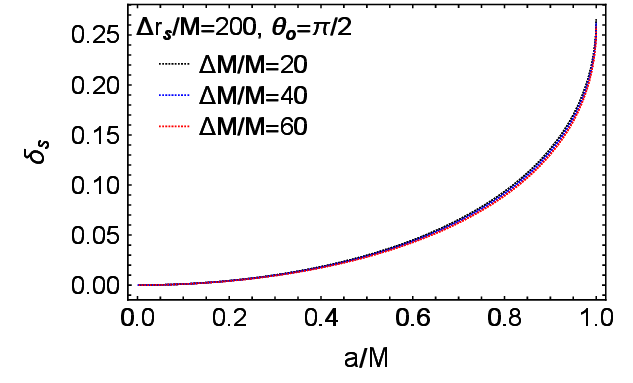}
\end{minipage}
\hspace{0.1cm}
\begin{minipage}{0.31\textwidth}
    \centering
    \includegraphics[scale=0.53]{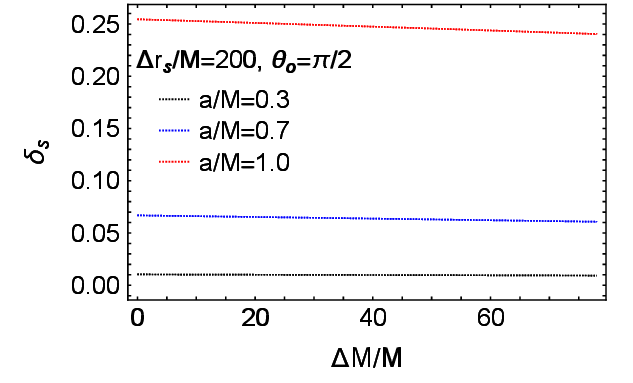}
\end{minipage}
\hspace{0.1cm}
\begin{minipage}{0.31\textwidth}
    \centering
    \includegraphics[scale=0.53]{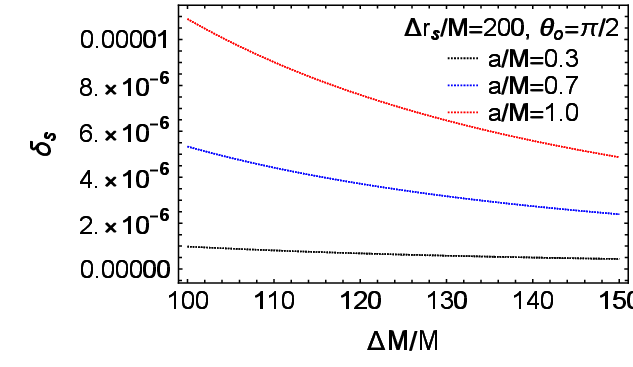}
\end{minipage}

\caption{\label{fig.7} The first three panels illustrate how the shadow radius varies with the spin parameter and the DM mass, while the last three panels show the dependence of the distortion parameter on these quantities.}
\end{figure}

Once the DM mass exceeds the critical threshold, as shown in the third column, the shadow radius grows by orders of magnitude and the influence of the spin becomes negligible. In this regime, the distortion parameter approaches zero, yielding an almost perfectly circular shadow. This highlights an interesting effect: sufficiently large DM masses can substantially alter the shadow of an axisymmetric BH, effectively restoring circular symmetry. The circularization observed at large DM mass has a physical origin. In the high mass branch shown in Fig. \ref{fig.7}, the horizon and the relevant photon region are displaced to much larger radial distances. In
particular, the critical photon orbits are located outside the DM shell, where the mass function is constant and the exterior geometry takes the Kerr form with a substantially increased total mass. Since
the spin parameter is kept fixed while the total gravitational mass increases, the relative importance of rotation and frame dragging becomes very small. The exterior geometry sampled by the critical photons consequently approaches the slowly rotating, and hence nearly Schwarzschild-like, regime. The difference between prograde and retrograde photon trajectories then becomes negligible compared with
the enlarged shadow radius, leading to an almost circular shadow. Thus, the circularization arises because rotational effects become progressively suppressed relative to the dominant total gravitational mass scale.

\subsection{Energy emission rate}
% In this section, we investigate the energy emission rate of rotating BH immersed in DM environment. The energy emission rate is a fundamental observable that encodes the thermodynamic and radiative properties of BH and is intrinsically linked to the underlying spacetime geometry. In the high-frequency (geometric-optics) limit, the absorption cross section is well approximated by the area of the BH shadow. Due to photon sphere constitutes a hypersurface of unstable null circular geodesics and can therefore be interpreted as an effective capture cross section of BH. Consequently, from the perspective of an observer at infinity, the BH shadow provides a direct probe of the high-energy absorption cross section. This correspondence provides a direct link between the optical properties of BHs and their radiative behavior \cite{Mashhoon:1973zz,Fabbri:1975sa,Page:1976df,Unruh:1976fm,Das:1996we}. In addition, the absorption cross section is widely recognized as a fundamental quantity characterizing the probability of an absorption process and scattering of various fields in the vicinity of BHs \cite{Jung:2004yh,Doran:2005vm,Crispino:2007qw}.  By analyzing the dependence of the energy emission rate on the black hole spin and the dark-matter mass, we aim to clarify the role of dark matter in shaping the emission behavior and to assess its potential impact on observable signatures of rotating black holes.

In this section, we investigate the energy emission rate of rotating BH immersed in DM environment. The energy emission rate is a fundamental observable that encodes both the thermodynamic and radiative properties of a BH and is intrinsically determined by the underlying spacetime geometry. In the high-frequency (geometric-optics) regime, the absorption cross section \(\sigma_{lim}\) can be accurately approximated by the area of BH shadow. This correspondence arises because the photon sphere forms a hypersurface of unstable null circular geodesics, which effectively acts as a capture cross section for incident radiation. As a result, from the viewpoint of an observer located at infinity, the BH shadow provides a direct probe of the high-energy absorption cross section. This connection establishes a clear link between the optical appearance of BHs and their radiative behavior \cite{Mashhoon:1973zz,Fabbri:1975sa,Page:1976df,Unruh:1976fm,Das:1996we,Jung:2004yh,Doran:2005vm,Crispino:2007qw}. 

For spherically symmetric BHs, the high-energy \(\sigma_{lim}\) which corresponds to the geometrical cross section of the photon sphere \cite{Mashhoon:1973zz,Decanini:2011xi}. Consequently, this correspondence can be naturally extended to rotating BHs. In Refs. \cite{Wei:2013kza,Hou:2018avu,Jusufi:2019nrn}, the absorption cross section is defined using \(R_{\rm sh}\) \((\sigma_{lim}\approx \pi R^2_{\rm sh})\). however, due to the irregular shape of the shadow, this definition may appear somewhat imprecise. Here, we adopt the average shadow radius to define the cross section, which is expressed as 

\be
\label{45}
\bar{R}=\frac{1}{2 \pi} \int_0^{2 \pi} \sqrt{\left(\alpha-\alpha_c\right)^2+\beta^2} d \gamma
\ee
where \((\alpha_c, 0)\) represents the geometric center of the shadow, and \(\gamma\) specifies the angle measured along the shadow boundary from the \(\alpha\) axis.
\be
\label{46}
\gamma=\arctan \left(\frac{\beta}{\alpha-\alpha_c}\right) .
\ee

Accordingly, we replace \(R_{\rm sh}\) with \(\bar{R}\), so that the energy emission rate can be written as

\be
\label{47}
\frac{d^2 E(\omega)}{d \omega d t}=\frac{2 \pi^3 \omega^3\bar{R}^2}{e^{\omega / T_{H}}-1},
\ee
here \(\omega\) denotes the photon frequency, while \(T_H\) represents the Hawking temperature of the BH evaluated at the outer horizon \(r_+\), whose location depends on both the DM parameter \(\Delta M\) and the spin parameter \(a\). Moreover, the Hawking temperature is given by \cite{PhysRevD.13.191}
\be
\label{48}
T_{H}=\frac{\kappa}{2\pi},
\ee
where \(\kappa\) denotes the surface gravity associated with the outer Killing horizon. It characterizes the gravitational acceleration at the BH horizon as measured relative to an observer at infinity. Since the Hawking temperature is directly determined by the surface gravity of the outer horizon, the corresponding Killing vector field generating this horizon can be expressed as
\be
\label{49}
\chi^\mu=(\partial_t)^\mu+\Omega_{\rm H}(\partial_\phi)^\mu,
\qquad
\Omega_{\rm H}=\frac{a}{r_+^2+a^2},
\ee
where \(\Omega_{\rm H}\) denotes the angular velocity of the BH. Next, using the definition of the surface gravity
\be
\label{50}
\kappa\equiv\sqrt{-\frac{1}{2}
(\nabla_\mu\chi_\nu)(\nabla^\mu\chi^\nu)
}\bigg|_{r=r_+},
\ee
we can obtain Hawking temperature \cite{PhysRevD.101.084023,Franzin:2022wai}
\be
\label{51}
T_H=\frac{\Delta'(r_+)}
{4\pi(r_+^2+a^2)}
=\frac{r_{+}^2 f^{\prime}\left(r_{+}\right)\left(r_{+}^2+a^2\right)+2 a^2 r_{+}\left(f\left(r_{+}\right)-1\right)}{4 \pi\left(r_{+}^2+a^2\right)^2}.
\ee

From the metric given in Eq. \eqref{23}, we readily obtain 
\be
\label{52}
T_H=-\frac{r_{+} \left(a^2+r^2_{+} \right) m'(r_{+} )+(a-r_{+} ) (a+r_{+} ) m(r_{+} )}{2 \pi  \left(a^2+r^2_{+} \right)^2}.
\ee

It is evident that, in the limit \( m(r) \to M \), the above expression consistently reproduces the Hawking temperature of the Kerr BH
\be
\label{53}
T_{\rm K e r r}=\frac{r_{+}^2-a^2}{4 \pi r_{+}\left(r_{+}^2+a^2\right)}.
\ee

Figure 8 illustrates the variation of energy emission rate as function of the particle frequency in the high-energy regime. For fixed DM mass, the energy emission rate is found to decrease monotonically with increasing spin parameter \(a\). This behavior can be understood from the Hawking temperature given in Eq. (48), which decreases as \(a\) grows. As the BH approaches the extremal limit, the corresponding reduction in temperature leads to a pronounced suppression of thermal radiation, and consequently to a weaker energy emission.

When the DM mass remains below the critical value, its influence on the emission rate is marginal, leading at most to a very slight enhancement. Once the DM mass exceeds the critical threshold, however, the Hawking temperature becomes extremely small, and the energy emission rate is strongly suppressed, effectively approaching zero. Such cases are therefore not displayed in the figure. Overall, these results demonstrate that DM can affect the energy emission properties of rotating BHs, although significant deviations arise only when the DM mass is large enough to induce substantial modifications of the spacetime geometry and its thermodynamic characteristics.

\begin{figure}[htbp]
\centering

% 第一行
\begin{minipage}{0.465\textwidth}
    \centering
    \includegraphics[width=\linewidth]{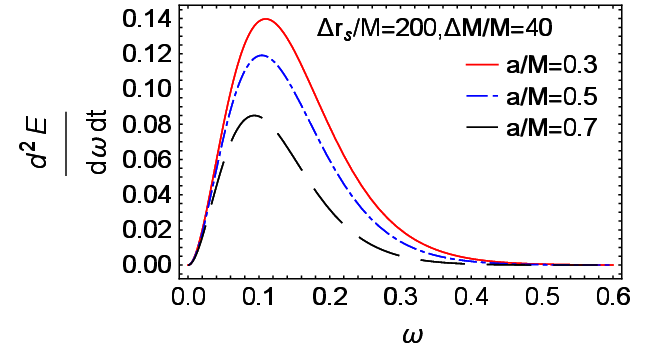}
\end{minipage}%
\hfill
\begin{minipage}{0.465\textwidth}
    \centering
    \includegraphics[width=\linewidth]{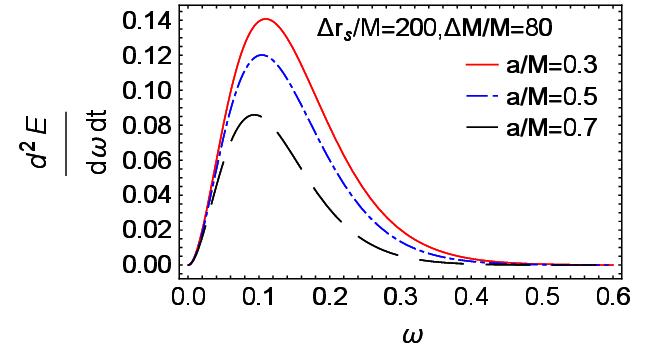}
\end{minipage}

\vspace{0.3cm}

% 第二行
\begin{minipage}{0.478\textwidth}
    \centering
    \includegraphics[width=\linewidth]{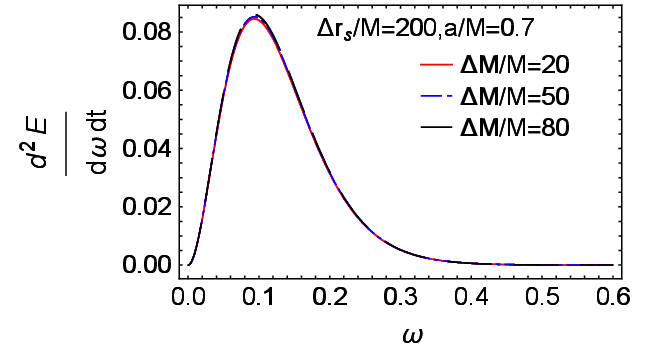}
\end{minipage}%
\hfill
\begin{minipage}{0.485\textwidth}
    \centering
    \includegraphics[width=\linewidth]{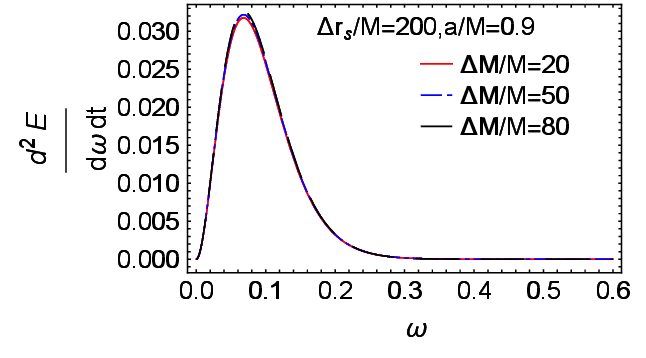}
\end{minipage}

\caption{\label{fig.8}The energy emission rate as a function of the particle frequency \( \omega \) for various values of spin \( a \) and DM mass  \( \Delta M \).
 }
\end{figure}

\subsection{Constraining DM Mass via the Shadow Diameter}

The shadow diameter measured by the EHT provides a direct means of constraining the amount of DM surrounding BH. In geometrized units, the observed dimensionless shadow diameter can be written as \(d_{\rm sh}^{\rm obs}=D\bar{\theta}/M\), where \(D\), and \(\bar{\theta}\) denote the source distance, and angular diameter, respectively. The corresponding theoretical prediction in our model is \(d_{\rm sh}^{\rm th}=2R_{\rm sh}\), with \(R_{\rm sh}\) defined by Eq. \eqref{43}. Current observations give \(d_{\rm sh}^{\rm obs}/M=11\pm1.5\) for M87* and \(d_{\rm sh}^{\rm obs}/M=9.77\pm0.67\) for Sgr A* \cite{Mizuno:2018lxz,EventHorizonTelescope:2021dqv,Vagnozzi:2022moj,EventHorizonTelescope:2022wkp,EventHorizonTelescope:2019dse,Bambi:2019tjh}. Therefore, the allowed range of the DM mass can be determined by requiring the theoretically predicted diameter to lie within the observational confidence intervals, while simultaneously restricting the parameters to the domain in which a regular event horizon exists.

In the present spacetime, the DM mass modifies the mass function and hence shifts the unstable photon region that determines the shadow boundary. When the DM mass is below its critical value, only a small fraction of the surrounding matter is enclosed by the relevant photon orbits, and the resulting correction to the shadow diameter remains weak. Once the DM mass becomes sufficiently large to generate the enlarged horizon branch, both the characteristic photon-orbit scale and \(R_{\rm sh}\) increase substantially. Consequently, excessively large values of the DM mass predict a shadow diameter exceeding the EHT compatible range and can therefore be excluded. 

\begin{figure*}[h]
\centering
\setlength{\tabcolsep}{0.9cm} 

\begin{tabular}{cc}
% ===== 第一张图 =====
\begin{minipage}[t]{0.4\textwidth}
\centering
\begin{overpic}[width=\textwidth]{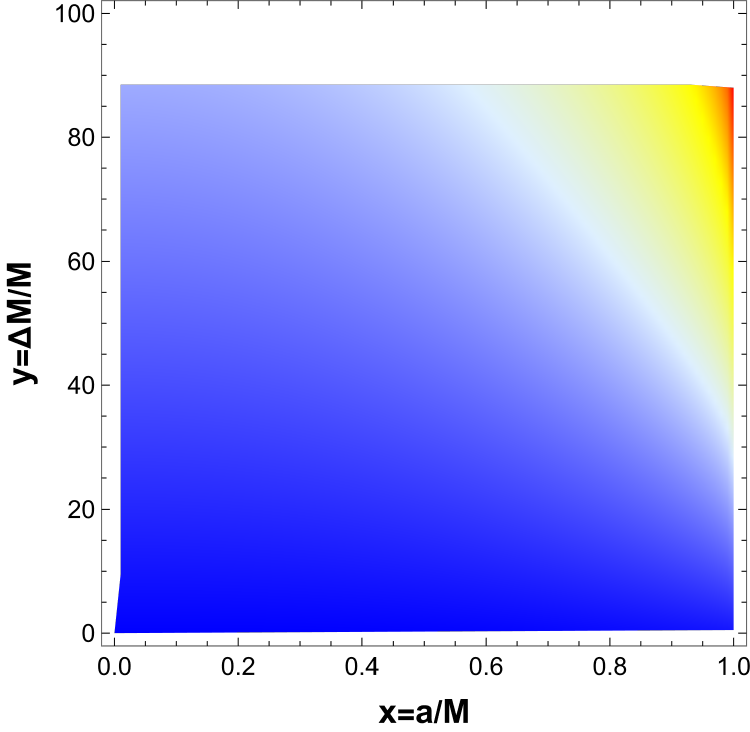}
    % 在主图中嵌入小条图 (位置通过坐标调整)
    \put(100,15){\includegraphics[width=0.20\textwidth]{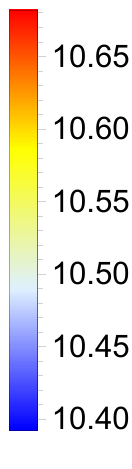}}
    \put(102,90){\color{black}\large $d_{sh}$} % 小图标签
\end{overpic}
\end{minipage}
&
% ===== 第二张图 =====
\begin{minipage}[t]{0.4\textwidth}
\centering
\begin{overpic}[width=\textwidth]{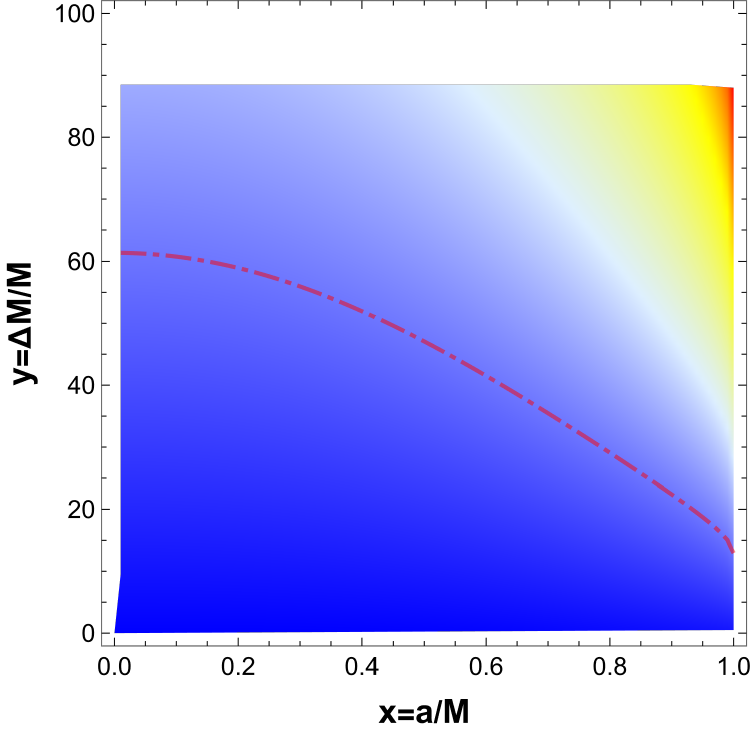}
    \put(100,15){\includegraphics[width=0.20\textwidth]{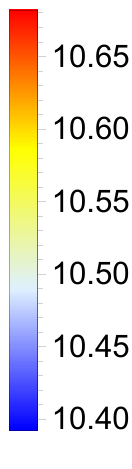}}
    \put(102,90){\color{black}\large $d_{sh}$}
\end{overpic}
\end{minipage}
\end{tabular}

\caption{\label{fig.9}{The figure presents contour plots of the shadow diameter over the theoretically allowed parameter space. The left panel corresponds to the contour plot for M87*. For Sgr A* (right panel), the red dashed lines indicate the \(1\sigma\) confidence region, corresponding to the observed shadow diameter \(d_{\rm sh}/M=10.44\).
}}
\end{figure*}

In Fig. \ref{fig.9}, we numerically present the constraints on the DM mass for M87* and Sgr A* within the theoretically allowed parameter space. The left panel corresponds to M87*, whereas the right panel shows the results for Sgr A*. The white region represents parameter values for which \(\Delta M/M\) exceeds the theoretical upper bound and is therefore excluded from our analysis. The colored regions show the contour maps of the shadow diameter. For M87*, the values of \(\Delta M/M\) throughout the entire colored region are compatible with the observational constraints, both within the \(1\sigma\) and \(2\sigma\) confidence intervals. For Sgr A*, the red dashed curve determines the upper bound on the DM mass at the \(1\sigma\) level. In other words, all parameter combinations below the red dashed curve are allowed within the \(1\sigma\) confidence region. If the \(2\sigma\) confidence interval is considered, the entire colored region becomes compatible with the observational constraint. The corresponding upper limits on the DM mass within the \(1\sigma\) confidence interval are summarized in Table 3. In particular, for \(a/M=0.9\), the DM mass is constrained to satisfy $\Delta M/M \leq 22.3853$.

\begin{table*}[h]
\centering
\caption{\label{Tab.3} The table presents the values of \( \Delta M/M \) at the \( 1\sigma \) and \( 2\sigma \) confidence levels for M87* and Sgr A*, with \( a/M \) held fixed.}
\begin{tabular}{|lllll|lllll|}
\hline
\hline
M87*$ \Delta M/M $& $1\sigma$ & $1\sigma$ & $2\sigma$ & $2\sigma$ &
Sgr A*$\Delta M/M$ & $1\sigma$ & $1\sigma$ & $2\sigma$ & $2\sigma$ \\
\hline
$a/M$ & Upper & Lower & Upper & Lower &
$a/M$ & Upper & Lower & Upper & Lower \\
\hline
0.1 & - & - & - & - & 0.1 & 60.7764 & - & - & - \\
\hline
0.2 &- & - & -& -   & 0.2 & 58.9389 & - & - & - \\
\hline
0.3 & - & - & - & - & 0.3 & 55.9538 & - & - & - \\
\hline
0.4 & - & - & - & - & 0.4 & 51.9416 & - & - & - \\
\hline
0.5 & - & - & - & - & 0.5 & 47.0644 & - & - & - \\
\hline
0.6 & - & - & - & - & 0.6 & 41.5103 & - & - & - \\
\hline
0.7 & - & - & - & - & 0.7 & 35.4661 & - & - & - \\
\hline
0.8 & - & - & - & - & 0.8 & 29.0877 & - & - & - \\
\hline
0.9 & - & - & - & - & 0.9 & 22.3853 & - & - & - \\
\hline
\end{tabular}
\end{table*}

\section{Conclusion}
In this work, we have successfully generalized the Schwarzschild BH surrounded by DM to the axisymmetric regime by employing the NJA method. By incorporating piecewise mass function into the Kerr metric, we constructed a rotating BH immersed in DM halo and verified that this solution satisfies the Einstein field equations. Our analysis of the spacetime structure reveals that the presence of DM significantly influences the event horizons and ergospheres. Specifically, when the DM mass exceeds a critical threshold, it induces a substantial expansion of these geometric boundaries, shifting their radial positions by an order of magnitude compared to the vacuum Kerr scenario.

Through the study of null geodesics and the derivation of celestial coordinates, we characterized the morphology of BH shadow and its associated observables. We observed that the shadow radius is positively correlated with the mass of the surrounding DM. As the mass increases, the shadow silhouette grows monotonically. Furthermore, while the spin of BH typically introduces a characteristic asymmetry, the surrounding DM exerts a strong circularizing effect. Consequently, for sufficiently large DM masses, the distortion parameter decreases markedly, causing the shadow to retain a nearly circular shape even in high spin regimes.

Finally, we investigated the energy emission rate based on the high-energy absorption cross section. Our results demonstrate that the emission rate is suppressed by both the BH spin and the presence of the DM halo. Notably, a large DM mass significantly reduces the Hawking temperature, leading to a drastic attenuation of the energy emission rate. Crucially, the substantial structural expansion resulting from high DM mass potentially violates current observational bounds. Next, we use the observed shadow diameters of M87* and Sgr A* to constrain the upper bound on the DM mass. For M87*, all theoretically allowed DM masses below the critical value are consistent with both the \(1\sigma\) and \(2\sigma\) observational constraints. For Sgr A*, the entire theoretically allowed DM mass range satisfies the \(2\sigma\) constraint, whereas an upper bound on the DM mass emerges at the \(1\sigma\) level. This upper bound decreases as \(a/M\) increases. In particular, for \(a/M=0.9\), the DM mass is constrained by $\Delta M/M \leq 22.3853$. This finding provides a potential observational criterion for testing the DM model considered here and offers a basis for future tests of gravity in the strong-field regime.

\begin{acknowledgments}
This study was supported by the Basic Research Program of Shanxi Province (Grant No. 202503021211204).
\end{acknowledgments}

\appendix

\section{Proof of the absence of the surface stress-energy tensor}

In this appendix, we demonstrate that the piecewise DM mass distribution $m(r)$ does not generate any surface stress-tensor at the boundaries.
A surface stress-energy tensor describes matter confined to an infinitesimally thin hypersurface and may arise when two geometries are matched improperly. Its presence is therefore determined by the junction conditions, rather than by the piecewise form of the matter distribution itself. To make this distinction explicit, consider a non-null interface \(\mathcal H\), with intrinsic coordinates \(y^a\) and tangent vectors \(e_a^{\mu}=\partial x^\mu/\partial y^a\). In a Gaussian normal coordinate \(\ell\), with the interface located at \(\ell=0\), the stress-energy distribution can be written as
\begin{equation}
\label{A1}
 T_{\m\n}= T^{\rm bulk}_{\m\n}+ S^{ab}e_a^\mu e_b^\nu\delta(\ell),
\end{equation}
where \(T^{\rm bulk}_{\m\n}\) are the regular bulk tensors, and \(S^{ab}\) is the surface stress-energy tensor. The last term represents a finite integrated source concentrated at the interface. Its presence is governed by the Darmois–Israel junction conditions. With
\begin{equation}
\label{A2}
h_{ab}=g_{\mu\nu}e_a^\mu e_b^\nu,
\qquad
K_{ab}=e_a^\mu e_b^\nu\nabla_\mu n_\nu,
\end{equation}
these conditions give
\begin{equation}
\label{A3}
8\pi S_{ab}
=-\epsilon\left([K_{ab}]-h_{ab}[K]\right),\qquad
\epsilon=n_\mu n^\mu=\pm1.
\end{equation}

Here \([Q]=Q_+-Q_-\), and the same orientation of the unit normal is used on both sides. In particular, continuity of the induced metric and extrinsic curvature implies \(S_{ab}=0\) \cite{Israel:1966rt,PhysRevD.105.024001,PhysRevD.43.1129}.

We now apply this criterion to the DM profile in Eq. \eqref{3}. Denote the inner and outer interfaces by
\begin{equation}
\label{A4}
R_1=r_s,\qquad R_2=r_s+\Delta r_s,
\end{equation}
and introduce
\begin{equation}
\label{A5}
x=\frac{r-r_s}{\Delta r_s}.
\end{equation}

Within the DM layer, Eq. \eqref{3} becomes
\begin{equation}
\label{A6}
m(r)=M+\Delta M(3x^2-2x^3),
\end{equation}
and hence
\begin{equation}
\label{A7}
m'(r)=\frac{6\Delta M}{\Delta r_s}x(1-x),
\qquad
m''(r)=\frac{6\Delta M}{(\Delta r_s)^2}(1-2x).
\end{equation}

The adjoining vacuum regions have constant mass functions \(M\) and \(M+\Delta M\), respectively. Consequently,
\begin{equation}
\label{A8}
m(R_i^-)=m(R_i^+)=M,
\qquad i=1,2.
\end{equation}
and
\begin{equation}
\label{A9}
m'(R_i^-)=m'(R_i^+)=0,
\qquad i=1,2.
\end{equation}

Thus, \([m]_{R_i}=0, [m']_{R_i}=0\). The profile is therefore \(C^1\) and piecewise smooth, although it is not globally \(C^2\). Additionally, the possible distributional terms can first be identified directly from the mass function. For a piecewise twice differentiable function, its second derivative in the distributional sense is
\begin{equation}
\label{A10}
D^2m=m''_{\mathrm{pw}}+\sum_{i=1}^{2}[m']_{R_i}\delta(r-R_i)+\sum_{i=1}^{2}[m]_{R_i}\delta'(r-R_i),
\end{equation}

where \(m''_{\mathrm{pw}}\) denotes the ordinary second derivative evaluated separately in each region. Both singular coefficients vanish for the present interpolation, giving $D^2m=m''_{\mathrm{pw}}$. In particular, the finite discontinuity of \(m''\) must be distinguished from a Dirac distribution in \(m''\). A Dirac contribution would require a discontinuity in \(m'\), which is absent here. The classical value assigned to \(m''\) at an isolated interface does not affect this distributional result.

This representation makes the matching particularly transparent. All dependence on the piecewise matter profile enters through \(m(r)\). The first derivatives of the metric consequently involve only \(m\) and \(m'\), both of which are continuous at the interfaces. Hence,
\begin{equation}
\label{A11}
[g_{\mu\nu}]_{R_i}=0,
\qquad
[\partial_\alpha g_{\mu\nu}]_{R_i}=0.
\end{equation}

This statement applies to the rotational cross terms as well as the diagonal components. Rotation changes the angular dependence and magnitude of the gravitational field, but introduces no discontinuity in its first derivatives for the adopted mass profile.
For a non-null interface, the induced metric and extrinsic curvature are constructed from the metric, its first derivatives, and the consistently chosen normal. Their one-sided limits therefore agree:
\begin{equation}
\label{A12}
[h_{ab}]_{R_i}=0,
\qquad
[K_{ab}]_{R_i}=0.
\end{equation}

Substitution into the junction condition yields
\begin{equation}
\label{A13}
S_{ab}^{(i)}=0,
\qquad i=1,2.
\end{equation}

Geometrically, the two adjoining regions induce the same intrinsic geometry and extrinsic curvature on their common boundary. Therefore, no surface stress-energy tensor is generated at the boundary.

\section{Proof that the Einstein Field Equations are satisfied}

In this appendix, to prove that the rotating BH incorporating DM, obtained via the NJA, satisfies the Einstein field equations, we first consider the field equations in the following form

\begin{equation}
\label{B1}
G_{\m\n}=8\pi T_{\m\n},
\end{equation}
where \(G_{\m\n}\) denotes the Einstein tensor, and \(T_{\m\n}\) denotes the energy–momentum tensor associated with DM. Through a series of calculations, we write the components of \(G_{\m\n}\) as
\be
\begin{aligned}
\label{B2}
G_{t t} & =\frac{2 \mathrm{Y}^{\prime}\left(a^4 \cos ^4 \theta-a^4 \cos ^2 \theta+a^2 r^2+r^4-2 \mathrm{Y} r^3\right)-a^2 r \sin ^2 \theta \mathrm{Y}^{\prime \prime}\Sigma}{\Sigma^3}, \\
G_{r r} & =-\frac{2 \mathrm{Y}^{\prime} r^2}{\Delta \Sigma} ,\\
G_{\theta \theta} & =-\frac{\mathrm{Y}^{\prime \prime} a^2 r \cos ^2 \theta+2 \mathrm{Y}^{\prime} a^2 \cos ^2 \theta+\mathrm{Y}^{\prime \prime} r^3}{\Sigma}, \\
G_{t \phi} & =\frac{a \sin ^2 \theta\left[r\left(a^2+r^2\right) \Sigma \mathrm{Y}^{\prime \prime}+2 \mathrm{Y}^{\prime}\left(\left(a^2+r^2\right) a^2 \cos ^2 \theta-a^2 r^2-r^3(r-2 \mathrm{Y})\right)\right]}{\Sigma^3}, \\
G_{\phi \phi} & =-\frac{\sin ^2 \theta\left[r\left(a^2+r^2\right)^2 \Sigma \mathrm{Y}^{\prime \prime}+2 a^2 \mathrm{Y}^{\prime}\left(\cos ^2 \theta\left(a^4+3 a^2 r^2+2 r^4-2 \mathrm{Y} r^3\right)-a^2 r^2-r^4+2 \mathrm{Y} r^3\right)\right]}{\Sigma^3}.
\end{aligned}
\ee

Moreover, we select an energy–momentum tensor constructed from an appropriately selected tetrad of the vector \(T^{\mu \nu}=e_a^\mu e_b^\n T^{a b}\), with its components satisfying \(T^{ab} = \text{diag}(\rho,p_{r},p_{\theta},p_{\phi})\). The components of energy–momentum tensor is then expressed in the orthonormal basis as follows

\be
\label{B3}
\begin{aligned}
\rho & =\frac{1}{8 \pi} e_t^\mu e_t^\nu G_{\mu \nu}, & & p_r=\frac{1}{8 \pi} e_r^\mu e_r^\nu G_{\mu \nu}, \\
p_\theta & =\frac{1}{8 \pi} e_\theta^\mu e_\theta^\nu G_{\mu \nu}, & & p_{\phi}=\frac{1}{8 \pi} e_{\phi}^\mu e_{\phi}^\nu G_{\mu \nu}.
\end{aligned}
\ee

Therefore, we need to find a basis that satisfies the field equations. One such orthogonal basis is the following choice

\be
\label{B4}
\begin{aligned}
e_t^\mu & =\frac{1}{\sqrt{\Sigma \Delta}}\left(r^2+a^2, 0,0, a\right), \\
e_r^\mu & =\sqrt{\frac{\Delta}{\Sigma}}(0,1,0,0), \\
e_\theta^\mu & =\frac{1}{\sqrt{\Sigma}}(0,0,1,0), \\
e_\phi^\mu & =-\frac{1}{\sqrt{\Sigma \sin ^2 \theta}}\left(a \sin ^2 \theta, 0,0,1\right).
\end{aligned}
\ee

Thus, for the Einstein field equations to be satisfied, the density and pressure of DM under this basis can be derived as

\be
\label{B5}
\begin{aligned}
\rho&=-p_r=\frac{ \mathrm{m}^{\prime}(r) r^2}{4 \pi \Sigma^2},\\ p_\theta&=p_\phi=-p_r-\frac{\mathrm{m}^{\prime \prime}(r) r+2 \mathrm{m}^{\prime}(r)}{8 \pi \Sigma},
\end{aligned}
\ee
where, $m^{\prime}(r)$ denotes $d m(r) / d r$. These results indicate that \eqref{23} is indeed a solution to Einstein field equations.

\bibliographystyle{unsrt}
\bibliography{Shadow}
\end{document}